\documentclass[letterpaper]{article}
\usepackage{aaai2026}
\usepackage{booktabs}
\usepackage{times}
\usepackage{helvet}
\usepackage{courier}
\usepackage[hyphens]{url}
\usepackage{graphicx}
\usepackage{natbib}
\usepackage[most]{tcolorbox}
\usepackage{caption}
\usepackage{xcolor}
\usepackage{comment}
\usepackage{algorithm}
\usepackage{algorithmic}
\usepackage{newfloat}
\usepackage{listings}
\usepackage{tabularx}

\usepackage{nameref}

\newcommand{\myparagraph}[1]{\textbf{#1.}}

\newcommand{\answerYes}[1]{\textcolor{blue}{#1}} 
 
\newcommand{\answerNA}[1]{\textcolor{gray}{#1}} 
 
\tcbset{
  myboxstyle/.style={
    colback=gray!10,    
    colframe=black,     
    boxrule=0.5pt,      
    arc=2mm,            
    left=2mm, right=2mm, top=1mm, bottom=1mm,
    breakable,          
  }
}

\newcommand{\newdatainfo}[1]{{#1}}
\newcommand{\added}[1]{{#1}}

\usepackage{amssymb}
\newcommand{\mybullet}{
    \vspace{0.4mm}
    \noindent  
    $\triangleright$
}

\DeclareCaptionStyle{ruled}{labelfont=normalfont,labelsep=colon,strut=off}
\floatstyle{ruled}
\newfloat{listing}{tb}{lst}{}
\floatname{listing}{Listing}

\title{Grok in the Wild: Characterizing the Roles and Uses \\ of Large Language Models on Social Media}

\author{
    Katelyn Xiaoying Mei\equalcontrib\textsuperscript{\rm 1}, 
    Robert Wolfe\equalcontrib\textsuperscript{\rm 1,2},
    Nicholas Weber\textsuperscript{\rm 1}, 
    Martin Saveski\textsuperscript{\rm 1}
}

\affiliations{
    \textsuperscript{\rm 1}University of Washington\\
    \textsuperscript{\rm 2}Rutgers University\\
    kmei@uw.edu, robert.wolfe@rutgers.edu, nmweber@uw.edu, msaveski@uw.edu
}

\begin{document}
\maketitle

\begin{abstract}
xAI’s large language model, Grok, is called by millions of people each week on the social media platform X. Prior work characterizing how large language models (LLMs) are used has focused on private, one-on-one interactions. Grok’s deployment on X represents a major departure from this setting, with interactions occurring in a public social space. In this paper, we systematically sample three months of interaction data to investigate how, when, and to what effect Grok is used on X. At the platform level, we find that Grok responds to 62\% of requests, that the majority (51\%) are in English, and that engagement is low, with half of Grok's responses receiving 20 or fewer views after 48 hours. We also inductively build a taxonomy of 10 roles that LLMs play in mediating social interactions and use these roles to analyze 41,735 interactions with Grok on X. We find that Grok most often serves as an information provider but, in contrast to LLM use in private one-on-one settings, also takes on roles related to dispute mediation, such as truth arbiter, advocate, and adversary. Finally, we characterize the population of users who prompted Grok and find that their self-expressed interests are closely related to the roles the model assumes in the corresponding interactions. Our findings provide an initial description of human-AI interactions on X, and a broader understanding of the diverse roles that LLMs may play in our online social spaces.
\end{abstract}


\section{Introduction}

During the last seven days of August 2025, we observed more than 7.1 million posts on X prompting an interaction with xAI's chatbot  @grok. People have long shared online spaces with social bots \cite{shao2018spread}, but the increasing sophistication of large language models and their integration into online social spaces represents a new paradigm of human-AI interaction. In experimental settings, LLMs have shown an impressive ability to empathize with users \cite{ovsyannikova2025third,rubin2025comparing} and even persuade humans on complex political issues \cite{costello2024durably,bai2025llm}.  Observational studies of chatbot logs, like WildChat \cite{zhao2024wildchat}, have shown that users disclose private and sometimes compromising information \cite{mireshghallah2024trust}, build emotional connections \cite{fang2025ai,zhang2025rise}, and over time reflect the opinions and communication style of LLM assistants~\cite{jakesch2023co}.

Frontier model providers, such as Microsoft, OpenAI, and Anthropic, have largely studied alignment and model use by analyzing anonymous chat log data. Findings from this work reflect the solitary nature of these products: In one-on-one interactions, LLMs are used as advanced search and computing interfaces for completing work, seeking advice, or offloading personal tasks \cite{tamkin2024clio}. Grok differs from other frontier models in that it was provided X's platform-specific data during both training and inference \cite{grok4modelcard}.  The model is intentionally designed and deployed to interact naturally in a social setting. This makes Grok's deployment to X a uniquely valuable site for observing whether the social impacts of LLMs anticipated by prior work (e.g., persuasion, emotional connection, and opinion influence) emerge at scale in naturalistic settings.

\added{Existing analysis of Grok on X has justifiably focused on its ethical shortcomings, sometimes prompting backlash severe enough to induce xAI to temporarily disable access to the model or certain features \cite{hagen2025mechahitler,taylor2025deletehitler}.}  But, amid public discussion of the \added{ethical and even legal perils} of Grok's use on X, an essential question with implications for \added{situating the model's limitations}, remains to be systematically addressed: How do millions of people interact, on an everyday basis, with one of the world's first intentionally social LLMs? 

In this study, we answer fundamental questions about how Grok is called, how it responds, and what roles it plays in everyday social life on X\footnote{Replication code and data are available in this repository: \\ \url{https://github.com/Mooniem/GrokInTheWild}}. We address three questions:

\begin{enumerate}
    \item \textbf{What are the overall usage patterns of Grok on X?} Specifically, how often is Grok called, and how often does it respond? In what languages is it most commonly called? How much engagement do Grok posts receive?
    \item \textbf{What kinds of interactions do users have with Grok, and what roles does the model play on the platform?} What are the characteristic patterns of Grok usage, and what do users expect from the model when they call it?
    \item \textbf{What characterizes user accounts that interact with Grok?} What are the interests and concerns of users who engage \added{with} Grok, and do they inform the model's roles? 
\end{enumerate}

\noindent To answer these questions, we collected $N$=\newdatainfo{41,735} user interactions with Grok consisting of \newdatainfo{142,895} posts from the official X API on an hourly basis \newdatainfo{between August 15, 2025 and November 17, 2025}, and queried the API for the profiles of users who engaged Grok, as well as for counts of Grok interactions \newdatainfo{over three months}. We adopted a mixed methods approach to analyzing this data, employing quantitative methods to characterize counts of Grok interactions and user profiles, and qualitative methods to describe categories of Grok usage and the social roles that users expect Grok to fulfill. We make the following contributions:

\begin{enumerate}
    \item A \textbf{quantitative analysis of Grok usage patterns} \newdatainfo{over three months}, finding that Grok replies to about \newdatainfo{62\%} of the posts prompting Grok for an interaction;
    that \newdatainfo{51\%} of posts prompting Grok are in English; and that engagement with Grok Reply posts is low, with half the posts receiving 20 or fewer views after 48 hours (Section \ref{sec:grok_usage_engagement}). 
    \item A description of the \textbf{categories of use} for which users employ Grok, finding that \newdatainfo{51\%} of Grok interactions involve information seeking, \newdatainfo{21.7\%} involve fact-checking, \newdatainfo{12.8\%} involve requests for opinions and advice, and \newdatainfo{12.4\%} are creative or generative in nature. We also find that context matters for the category of use, as fact-checking is more common deep in conversation threads, while requests for opinions and advice or creative interactions are more common at the beginning of a thread (Section \ref{sec:results-use-and-roles}). 
    \item A description of \textbf{social roles assumed by Grok}, taking user expectations, the context of interaction, and broader function on X into consideration. We observe emergent social roles for LLMs in our data, including Adversary (characterized by taking a position opposed to a user), Truth Arbiter (by deciding questions contested by multiple users), and Platform Insider (by providing detailed information about X and its users) (Section \ref{sec:results-use-and-roles}).
    \item Analysis of \textbf{user accounts that engage with Grok on X}, finding that Grok users are relatively active on the X platform, with more than 75\% of users having posted at least 1,000 times, and more than 75\% of users having an account at least \newdatainfo{1.6} years old. By fitting topic models to user bios and \added{using an LLM to classify the roles Grok plays in interaction with those users}, we find that user interests and concerns (\textit{e.g.}, political leanings and tech culture) inform the roles Grok plays on X (Section \ref{sec:user_analysis}).
\end{enumerate}

\noindent Our findings offer an initial characterization of what may ultimately become a mainstream form of social interaction with conversational AI, on social media platforms originally designed for human-to-human communication. This study was approved by the University of Washington's IRB.

\section{Background and Related Work}

\myparagraph{Grok} xAI released Grok-1, a 314-billion-parameter model, in November 2023 \cite{xai2023announcinggrok}. The current version (Grok-4), \added{as of the time of this study} incorporates reasoning and multimodal capabilities, significantly advancing its information retrieval, analysis, and generation. In early August 2025 (the beginning of data collection of this study), xAI made the ``@grok'' feature available to all users on X, but unpaid X users were subject to more restrictive usage limits than paid users (exact limits on usage were not disclosed) \cite{wright2025free}. Users can request a text reply from Grok by mentioning ``@grok'' in a post, or request an explanation by clicking the button on the top right of a post (which opens a chat interface on the user's screen). In this work, we focus solely on the public interactions with Grok in which the model replies to @grok mentions. Previous studies examining these interactions have analyzed Grok's effectiveness in fact-checking information~\cite{Renault2026} and its use in correcting misinformation while avoiding personal attacks~\cite{caramancion2026using}, tasks that, as we show in Section~\ref{sec:results-use-and-roles}, are often reflected in the roles played by Grok on X (\textit{Truth Arbiter} and \textit{Advocate}).

\noindent \myparagraph{Bots on Social Media} Grok is among the first LLM-driven bots deployed at scale by a social media platform, but social bots have long existed on X and other social spaces~\cite{varol2017online}. Over the last decade, bots on social media have been tied to the spread of misinformation and conspiracy theories, and have drawn significant criticism for their potential interference with political elections ~\cite{ferrara2020types}. Given recent advances in LLMs, researchers have sought to use more powerful general-purpose models to counter misinformation \cite{zhou2024correcting,wu2025seeing}. However, concerns about LLMs as agents on social media remain, and research has started to examine how these increasingly anthropomorphic bots would interact with users. \citet{yang2023anatomy} identified a malicious bot network on X that appeared to use ChatGPT to generate realistic synthetic content. \citet{radivojevic2024llms} found that participants only correctly identify LLMs about 42\% of the time when asked to do so in a simulated social environment including humans and persona-prompted models. Most recently, \citet{moller2025impact} studied LLMs in a realistic social media environment, finding LLM tools increase the amount of content online and foster user engagement with it, but also negatively impact conversation quality and authenticity.

\noindent \myparagraph{Human-AI Interaction in the Era of LLMs}
Because LLMs exhibit general-purpose capabilities \cite{eloundou2024gpts}, including engaging in conceivably any form of conversation, recent studies attempt to characterize the ways that people actually use these technologies in the wild. Studies of public conversation logs with LLMs and proprietary datasets have identified a diverse range of user interactions~\cite{tamkin2024clio,mireshghallah2024trust}. Most of these interactions are productivity-oriented, supporting tasks across domains including programming, academic writing and education, content creation, and communication assistance. In addition to productivity-driven tasks, studies find that users discuss personal topics with LLMs and seek interpersonal advice and emotional support from them~\cite{fang2025ai,ammari2025studentsreallyusechatgpt,karnam2026bowling}. Recent work contends that LLMs could improve communication on X, including by intervening early to counter misinformation~\cite{li2025scaling,de2025supernotes}. \citet{argyle2023leveraging} find that employing LLMs prompted to make evidence-based recommendations can improve perception of reciprocity, tone, and conversation quality. \citet{tessler2024ai} find LLMs can help humans arrive at common ground by producing mediating statements that generate agreement from divided groups. We build on this rapidly maturing literature by examining how an LLM is used on a public social media platform, where discourse is often one-to-many and the model may be called by users at any time. 

\section{Data}

\noindent \myparagraph{Grok Usage Volume (RQ1)} 
To investigate usage patterns of Grok, we collected the number of posts in which users mentioned ``@grok’’ and the number of times Grok replied daily between August 15 and November 17, 2025. We collected these counts both overall and for English-only posts. To estimate Grok usage in different languages, we also collected counts for the most commonly used languages after English (estimated based on an initial sample) over one week, from September 2 to 8, 2025. We provide more details about the queries and API endpoints used in Appendix Section \ref{app:data}.

\begin{figure}
    \centering
    \includegraphics[width=\linewidth]{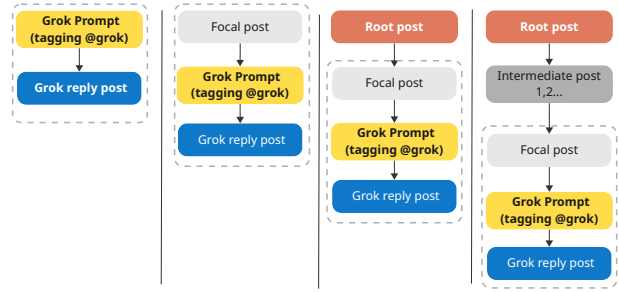}
    \caption{Four forms of focal post chains in our data.} 
    \label{fig:focal-tweet-structure}
\end{figure}

\noindent \myparagraph{Grok Interaction Posts (RQ2)} To describe the categories of use and social roles played by Grok on X, we collected \newdatainfo{41,735} interactions with Grok, consisting of \newdatainfo{142,895} posts from conversation threads on X that allowed us to study interactions between users and Grok. 
Figure \ref{fig:focal-tweet-structure} visualizes the four forms the chains could take. Each chain included a reply from Grok to a user (the \textbf{Grok Reply} post); the post from the user that prompted Grok to reply (the \textbf{Grok Prompt} post); the immediate parent (if it exists) of the Grok Prompt post (the \textbf{Focal} post); and the root post of the conversation (the \textbf{Conversation Root} post). The Grok Prompt and the Grok Reply constitute the primary interaction between Grok and user; the Focal and Conversation Root posts provide essential context for understanding the conversation. Each focal post chain thus included at least two and no more than four posts. Note that Conversation Root and Grok Prompt refer to the same post in some conversations (i.e., when Grok Prompt begins a thread, such that post ID matches conversation ID).
We collected a sample of interactions with Grok over a three-month period, from August 15 to November 17, 2025.
In keeping with the post collection limits of our developer account, we collected 500 conversations per day. To ensure we collected a sample of posts temporally representative of Grok usage, \added{we retrieved posts per hour in proportion to the number of Grok prompt posts at a given hour of the day observed in the week prior to data collection (Appendix Table~\ref{tab:convhour}). Appendix Section \ref{app:data} provides additional details of our sampling procedure and how we reconstruct the chain conversations using Grok Reply and Grok Prompt posts.} 

As described in Section \ref{sec:method}, our analysis also considered engagement metrics (views, likes, retweets, and replies) for Grok Reply posts, which we received as part of the response payload from our request to the \texttt{recent-search} endpoint. However, because we retrieved these posts just after they were posted, the engagement metrics received were usually very close to 0, and unlikely to reflect actual engagement with the model. We thus used the \texttt{tweet-lookup} endpoint to re-retrieve Grok Reply posts and associated engagement metrics 48 hours after first retrieving the posts. We successfully re-retrieved \newdatainfo{35,375 posts (85\%)}, as some posts had been deleted. Manually spot-checking deleted posts, we found that in most cases, the Grok reply was deleted when the Grok Prompt post had also been deleted. 

\noindent \myparagraph{User Data (RQ3)} To understand how user characteristics inform Grok usage patterns and social roles, we collected profile data using the \texttt{user-lookup} endpoint for the users who authored a Grok Prompt post in our dataset. We collected \newdatainfo{31,111} user profiles, lower than the number of Grok Prompt posts in our data due to the suspension of some accounts, and the presence of multiple posts by some users.

\section{Methods}\label{sec:method}

\subsection{Grok Interaction Analysis (RQ1)}

Using data from the \texttt{recent-tweet-counts} endpoint, we computed the following statistics about Grok use on X:

\begin{itemize}
    \item \textbf{Aggregate Counts Over Three \newdatainfo{Months}}: Counts of number of user mentions of Grok (requests for a reply) and Grok replies to posts, in English and overall on X.
    \item \textbf{One-Week Use in 11 X Languages}: Counts of Grok prompts and replies in 11 languages with the most Grok use after English. 
\end{itemize}

\noindent Based on data from the dataset of \newdatainfo{41,735} chains collected for RQ2, we computed the following additional statistics:

\begin{itemize}
    \item \textbf{Grok Reply Engagement}: The distribution of views, likes, retweets, and replies to the Grok Reply posts in our dataset, 48 hours after the Reply was initially collected.
    \item \textbf{Conversation Root Engagement}: The distribution of views, likes, retweets, and replies to the Conversation Root posts in our data, taken when the post was collected. 
\end{itemize}
\vspace{-2mm}
\subsection{Uses and Roles of Grok on X (RQ2)}
\vspace{-1mm}
To understand the roles assumed by Grok on X, we draw on Role theory, a psychological framework asserting that \textit{roles} are defined by ``characteristic'' social patterns, induced by the expectations of other individuals and groups, the immediate context of a social interaction, and the broader function of social behavior within a social system \cite{biddle2013role}. Roles are, importantly, not representative of \textit{all} of an individual's interactions; rather, they are dependent on context, such that one individual might alternately take on roles like manager and mother \cite{eagly2012social}. In accordance with role theory, understanding Grok's roles on X requires two steps. First, we describe the model's uses, including their frequency, which we accomplish via an initial content analysis and subsequent LLM-based classification across our dataset. Second, we characterize roles as patterns of interaction based on the \textit{expectations} users have for the model, the \textit{context} in which interactions occur, and, to the extent possible, the \textit{function} Grok interactions play in the social ecosystem of the X platform, necessitating a deeper contextual analysis of a larger sample representative of the use categories in our data.

\noindent \textbf{Grok Uses: Taxonomy Construction.} 
Four authors met weekly over three weeks to discuss a sample of the data collected. Before each meeting, the authors open-coded a sample of 200 conversations collected the preceding week. In the first meeting, the authors employed an affinity diagramming approach wherein they individually wrote observations about the data on post-it notes in a shared Miro board. The authors then collectively reviewed the notes and constructed a set of interaction categories to organize the data. In subsequent meetings, the authors discussed newly collected data, adding and revising categories to accommodate new observations. At the end of the third meeting, the authors agreed on ten categories of user interactions with Grok, as well as an Other category for notable but atypical interactions.

\noindent \textbf{Grok Uses: LLM Classification in the Full Dataset.} 
After arriving at a set of categories, we sought to apply them to the full dataset by prompting an LLM to classify the \newdatainfo{41,735} focal post chains collected to answer RQ2. To do so, one author created a set of instructions for an LLM to use in classifying each focal post chain, which a second author then reviewed and provided feedback on. These two authors then coded 100 focal post chains according to the 11 categories of use. After coding, the authors jointly reviewed their coding, identifying differences and resolving them through discussion. Once the codes had been finalized, the authors prompted three LLMs (GPT-5-Mini, Gemini-2.5-Flash, and Gemini-2.5-Pro) to classify these 100 focal post chains according to the categories. They then computed Cohen's $\kappa$ between the human-applied codes and the codes applied by each of the models. The highest IRR was achieved by Gemini-2.5-Pro, with $\kappa$ = .60, indicating substantial agreement. The authors then further refined the prompt for the model to handle edge cases leading to model confusion for semantically related categories. After updating the prompt, Gemini-2.5-Pro achieved $\kappa$ = .70. \added{We found that Gemini sometimes misinterpreted user debates with Grok as information seeking or fact-checking, especially when the chain involved the exchange of (disputed) factual information. Gemini also classified many chains as Other when the user’s instruction to Grok was underspecified, whereas we were able to assign a label by reviewing the thread in its entirety on the X platform.} We thus used Gemini-2.5-Pro with the updated prompt to classify the full dataset. We include prompts in Appendix Section~\ref{app:llmprompt}.

\noindent \textbf{Grok Roles: Taxonomy Construction.} 
We next sought to describe the roles that Grok plays on the X platform, based on a larger sample of 1,002 focal post chains stratified according to the frequency of each usage category, resulting in a minimum of 30 chains for the smallest category. Full counts by category are included in the Appendix Table~\ref{tab:sample_weight}. Two authors then open-coded these posts, situating them in their full conversation threads, taking note of 1) the user and Grok behaviors exhibited; 2) the expectations the user appeared to have for the model; 3) the social context of the conversation thread in which the interaction occurred; and 4) where identifiable, the social function that the user's interaction with the model appeared to play on X. The authors then met and discussed their observations, using affinity diagramming to group observations into roles, and assigning representative posts and quotes from posts to illustrate the~role. 

\noindent \textbf{Grok Roles: Placing Roles in Context of Prior Work.} 
After defining roles through contextual analysis, we sought to situate those roles in 1) the prior work on LLM uses and roles; and 2) the literature differentiating the function of social roles. For 1), we collectively reviewed 17 prior studies on LLM uses and roles (Appendix Section \ref{app:llm_usage_studies}). We then collated roles and uses observed in prior work, and identified the roles produced from this study that appear to reflect distinct patterns of usage compared to prior studies. For 2), we plot roles on two axes. The first axis considers roles as supportive vs. constraining of user behavior on X, drawing on the supportive-autonomy vs. controlling social environments described in psychological theories of motivation \cite{oliver2008effects}. The second considers the function of roles as egocentric (concerned with the individual user) vs. allocentric (concerned with the broader social environment), drawing from the work of \citet{karahanna2018needs} on how social media platforms meet users' psychological needs.

\noindent \added{\textbf{Grok Roles: LLM Classification in Full Dataset.} We used an LLM to assign a single role to each Grok interaction chain. To develop the prompt, two pairs of authors independently coded 50 randomly sampled chains each with primary and secondary roles, achieving agreement of $\kappa=.78$ and $\kappa=.74$. Common disagreements involved interpreting tone or rhetorical device, for instance in distinguishing information requests (Oracle) from seeking ammunition for disputes (Advocate). After resolving differences, this set ($n=100$) served as the validation data. Two authors then coded 99 additional interactions ($\kappa=.72$), to act as a test set. We then experimented with prompting strategies, achieving the best performance on the validation set ($\kappa=.86$) using a combination of few-shot prompting (ten labeled examples) with a two-stage ``prompt chaining'' classification: the first stage assigns post chains to broad categories (neutral information seeking, position-driven disagreements, platform-specific queries, or other), and the second assigns a specific role. Using Gemini-Flash-3-Preview,\footnote{\added{We selected Gemini-Flash-3-Preview based on its higher agreement ($\kappa=.77$) compared to Gemini-2.5-pro ($\kappa=.68$) on the test set. Because Flash-3-preview was released after the Grok Uses classification, it was not included in that stage.}} we achieved $\kappa=.76$ on the role label against our test set (prompt included in code release). We reported the distribution of roles in our dataset based on classification using this prompt.}

\noindent\myparagraph{Interactions Topic Model} Because social roles also depend significantly on the social context in which they are performed, we also fit a topic model to describe the snapshot of the X ecosystem present in our dataset. We computed embeddings of unique Conversation Root posts with Google Gemini using the \texttt{gemini-embedding-001} endpoint, setting task type to \texttt{clustering}. We then used a BERTopic pipeline to fit a topic model \cite{grootendorst2022bertopic}, setting minimum cluster size to \newdatainfo{100} examples. We reduced dimensionality using UMAP with \newdatainfo{40} neighbors to preserve non-linear relationships among documents \cite{mcinnes2018umap}. \added{These parameters were selected after several iterations of experimentation with the goal of obtaining meaningful cluster representations. To provide more transparency regarding how our choice of parameters affects the results, we present the results of two topic models: higher-level topics (Figure~\ref{fig:content_topic_map_primary}) and more granular topics (Figure~\ref{fig:content_topic_model_secondary}).} To generate more interpretable labels, we used OpenAI's GPT-4o-mini as the representation model, allowing it to review the content of representative posts when generating a topic label.

\vspace{-2mm}
\subsection{User Analysis (RQ3)}

We computed the following statistics based on profile information obtained from the \texttt{user-lookup} endpoint for authors of Grok Prompt posts in our dataset:

\begin{itemize}
    \item \textbf{User Activity:} We describe the distribution of user activity metrics, including number of posts and likes. 
    \item \textbf{User Network Size:} We describe the number of followers and the number of followees. 
    \item \textbf{User Account Age:} We compute the distribution of account ages in days since account creation.
    \item \textbf{X Verification}: We compute the percentage of users who are verified on the X platform. 
\end{itemize}

\noindent \myparagraph{User Topic Model} We created an additional topic model to describe account bios (users' own description of their accounts), using the same topic modeling approach as described above for X conversations (i.e., Gemini embeddings and a BERTopic pipeline). Because the user dataset was smaller and most descriptions were shorter than the posts used for RQ2, we set the minimum cluster size to \newdatainfo{50} examples, and set neighbors to \newdatainfo{20} for dimension reduction.

\noindent \added{\myparagraph{Computing Proportion of Grok Roles by User Cluster} We then sought to understand whether Grok assumed specific roles when called by the different clusters of users identified by our topic model. To do so, we computed the proportion of times that Grok assumed each role in a given user cluster, using the LLM classification of Grok roles for each post chain in our dataset (i.e., the proportions reported across all roles for a given cluster).}

\begin{figure}[!t]
    \centering    \includegraphics[width=\linewidth]{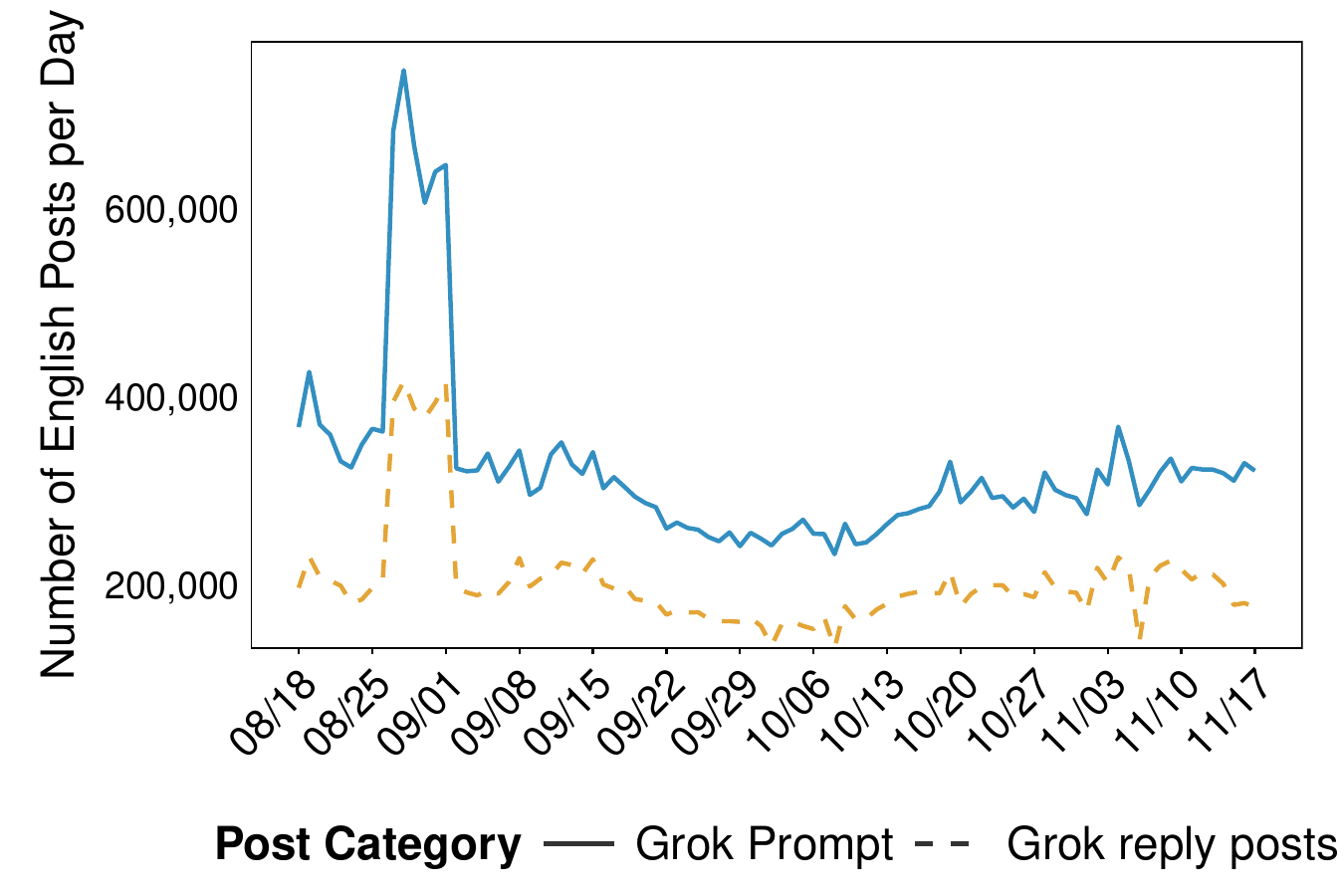}
    \caption{Time series of English-language posts mentioning ``@grok'' (Grok prompt) and Grok reply posts per day during our data collection period (August 18–\newdatainfo{November 17, 2025}). About 62\% of ``@grok'' mention posts receive a Grok reply, 
    indicating that user interest exceeds actual interaction.} 
    \label{fig:daily_traffic_grok_usage_english}
\end{figure}
\section{Results}

\subsection{Grok Usage Analysis (RQ1)}\label{sec:grok_usage_engagement}
\paragraph{Overall Usage of Grok on X.} 
%
We observe that about 62\% (ratio = 0.623, 95\%~CI~=~[0.620,~0.625]) of the hourly count of posts mentioning ``@grok'' (median~=~21,048, mean~=~23,194) receive a Grok Reply (median~=~14,213, mean~=~13,088), and that the hourly fluctuation in Grok Reply posts closely tracks that of Grok Prompt posts. 

A little over half of posts mentioning ``@grok'' are English-language posts (median~=~13,563, mean~=~12,514), a ratio also observed for Grok Reply posts (in English, median~=~7,896, mean~=~8,499). Around 65\% (95\%~CI~=~[0.634,~0.640]) of English-language posts mentioning ``@grok'' receive a Grok Reply, indicating that Grok replies to posts in English at a slightly higher rate than overall. We found that \newdatainfo{5\%} of Grok Prompt posts included media (image, GIF, or video preview), compared to only \newdatainfo{0.1\%} of posts authored by Grok. We plot the volume of English-language Grok Prompts and replies during our data collection period in Figure~\ref{fig:daily_traffic_grok_usage_english}, observing a relatively constant amount of engagement, with the exception of the sharp increase in usage between August 27 and September 1. We cannot say for certain what accounts for this increase, but note that the Grok account posted a marketing video on August 27 that received more than 78 million views, potentially inducing a temporary spike in usage.

As shown in Figure~\ref{fig:language_distribution}, the most common languages for ``@grok'' prompt posts were \newdatainfo{English (51\%), Spanish (6\%),  Japanese (4\%), and Portuguese (4\%). We observe frequent Grok usage in French (3\%), Arabic (3\%),  Turkish (3\%) and Hindi (2\%).}

\begin{table}[!t]
\centering
\footnotesize
\begin{tabular}{lr}
\toprule
\textbf{Type of Interactions} & \textbf{Count\ \ \ (\%)\ \ } \\
\midrule
Root $\to$ ... $\to$ Grok Call $\to$ Grok Reply & \newdatainfo{20,672 (49.6\%) }\\
Root $\to$ Grok Call $\to$ Grok Reply & \newdatainfo{15,272 (36.7\%) }\\
Root $\to$ Focal $\to$ Grok Call $\to$ Grok Reply & \newdatainfo{3,067 (\ \ 7.4\%) }\\
Root = Grok Call & \newdatainfo{2,648 (\ \ 6.4\%) }\\
\bottomrule
\end{tabular}
\caption{Distribution of Focal Post Chains}
\label{tab:focal_chain_types}
\end{table}

\noindent \myparagraph{Forms of Interaction with Grok}
We describe the frequency of four types of interactions with Grok. As shown in Figure~\ref{fig:focal-tweet-structure}, users may choose to mention Grok in a Conversation Root post, in a direct reply to a Conversation Root, or in a reply deeper in the conversation thread. Table~\ref{tab:focal_chain_types} shows that, of our sample of \newdatainfo{41,735} conversation chains, most (\newdatainfo{49.6\%}) mention Grok deeper in the conversation thread, indicating that users often mention Grok in conversations that have already generated some discourse\added{, including other interactions with Grok}. The second most common type of interaction (\newdatainfo{36.7\%}) mentions Grok in a reply to a Conversation Root. Only \newdatainfo{7.4\%} of chains mentioned Grok in a Conversation Root post, a ratio observed in part because we sample Grok replies not conversations, but nonetheless indicating that the typical Grok interaction occurs deeper within a conversation.

\begin{figure}[!t]
    \centering    
    \includegraphics[width=0.94\linewidth]{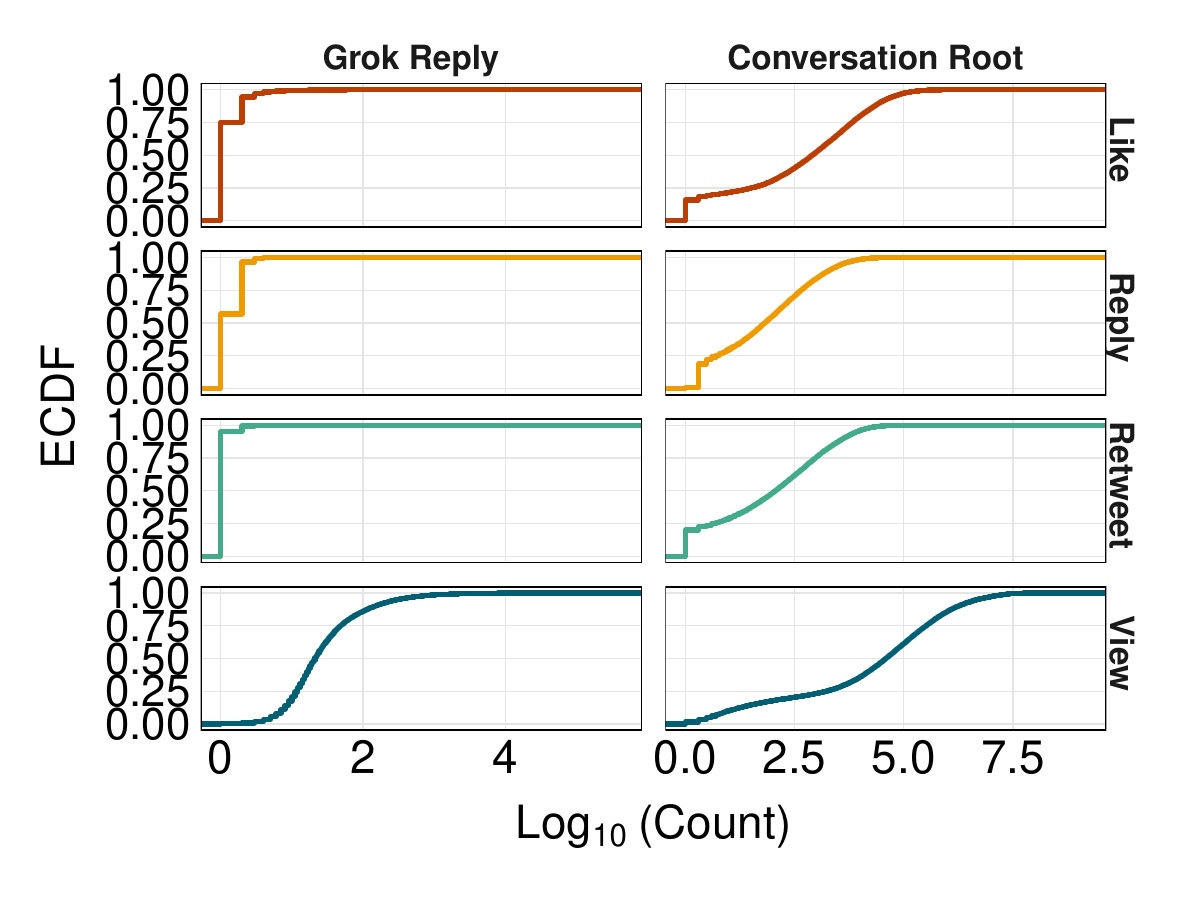}
    \caption{
    Cumulative distribution of engagement metrics (likes, replies, retweets, and views) for Grok Replies vs.\ Conversations Root posts. Grok replies receive significantly less engagement. 
    }
    \label{fig:ecdf_engagement}
\end{figure}

\paragraph{User Engagement: Grok Replies vs. Conversation Roots.} Most of the Conversation Root posts in our sample attracted substantial attention, receiving a median of \newdatainfo{39,304} views and a range from 0 to \newdatainfo{1.52 billion} views, as well as a high number of likes (median~=~805, mean~=~12,928). Compared with Conversation Root posts, Grok Reply posts received lower engagement across all metrics (views, likes, replies, retweets), including fewer views (at 48 hours: median~=~20, mean~=~147) and fewer likes (at 48 hours: median~=~0, mean~=~0.78). Moreover, 56.9\% of Grok Reply posts received 0 replies; 75.2\% 0 likes; and \newdatainfo{95.2\%} 0~retweets. To make the difference clear, we plot the cumulative distribution of log counts of engagement for both Conversation Root posts and Grok Reply posts in Figure~\ref{fig:ecdf_engagement}. \added{We also compared the distributions of these engagement metrics using two-sample Kolmogorov-Smirnov tests and found significant differences between Grok replies and root posts across all engagement metrics (all $p$-values $< 0.001$, see Appendix Table~\ref{tab:engagement_comparison}).}
\added{These differences in engagement are partially explained by network position and posting time. Grok calls are 1) posted by users with significantly fewer followers than those who posted the root posts ($t$-test on logged counts, $t(37,864) = -41.05$, $p < 0.001$); 2) posted slightly more often during less active times of the day in the U.S. ($\chi^2 (23) =309.69$, p$< 0.001$); and 3) posted a median of 7.8 hours after the root posts, when most of the attention on the post thread has likely passed.} 


%
Overall, our results suggest that X users are much less likely to observe and interact with the Grok posts that reply within a conversation than with root posts that appear in their X feed, and that Grok engagement is limited even in absolute terms, with half of the posts being viewed 20 or fewer times.

\subsection{Grok Use Categories and Social Roles (RQ2)} \label{sec:results-use-and-roles}
\begin{figure}
    \centering
    \includegraphics[width=\linewidth]{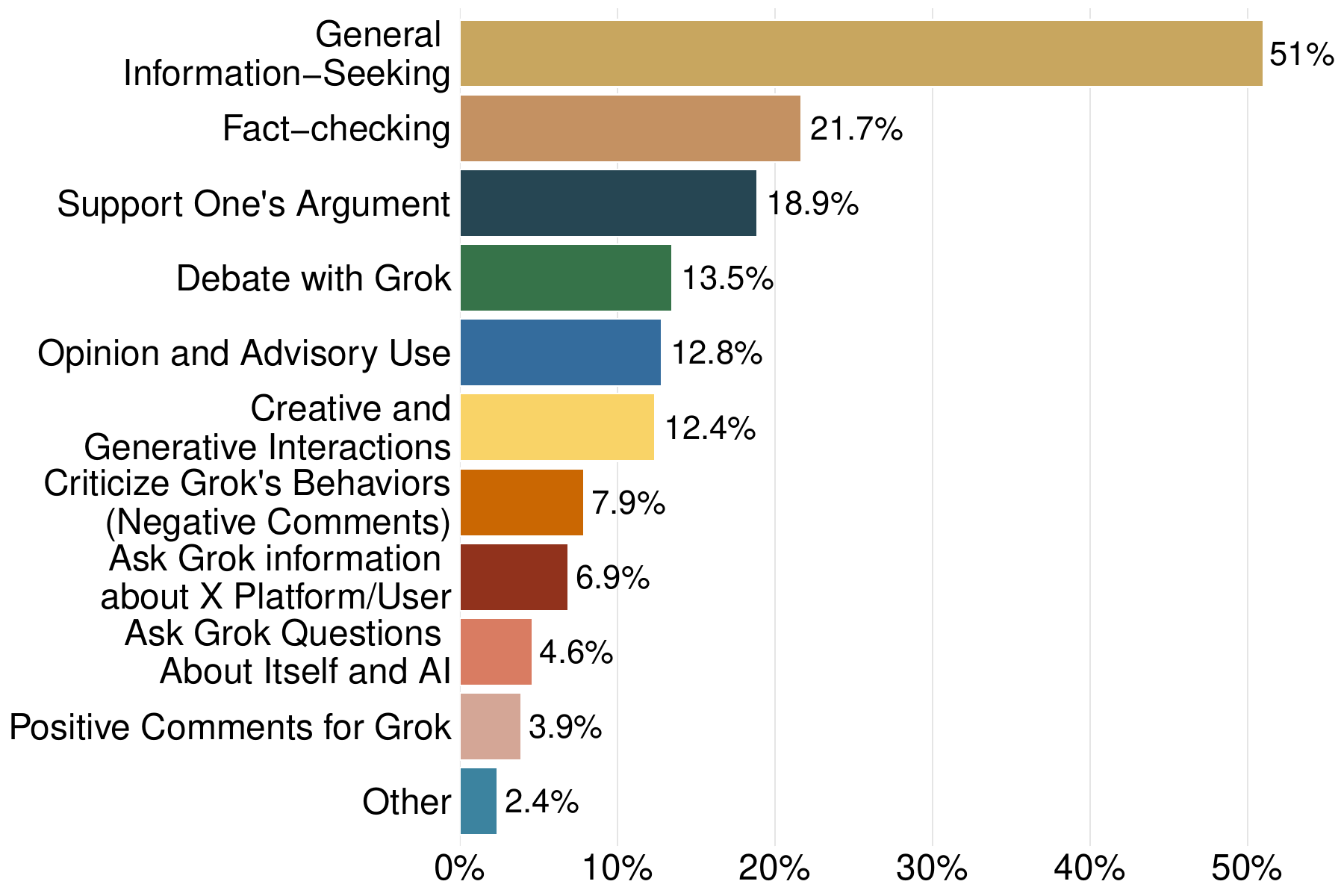}
    \caption{Distribution of categories of Grok use. Each interaction can be assigned to more than one category. Information seeking and fact-checking were the most common uses.} \label{fig:overall_classification_distribution}
\end{figure}

\noindent \textbf{How do people use Grok on X?} Figure~\ref{fig:overall_classification_distribution} illustrates the most common kinds of interactions users have with Grok, based on LLM classification of our dataset according to the 11 use categories we arrived at during qualitative analysis. Consistent with X's positioning of Grok, \textit{General Information-Seeking} (obtaining information to satisfy one's needs or curiosity) is the most common interaction type (51.0\%), followed by \textit{Fact-Checking} (determining the truth value of a claim or theory) (\newdatainfo{21.7\%}). Users also leveraged Grok in discussions and argumentation, with \newdatainfo{18.9\%} of interactions falling into the category \textit{Support One's Argument}, and \newdatainfo{13.5\%} into \textit{Debate with Grok}. Users also engaged Grok to ask questions about itself (\textit{Asking Grok Questions About Itself and About AI}, \newdatainfo{6.9\%}), to offer feedback in the form of \textit{Positive Comments for Grok} (\newdatainfo{3.9\%}) or to \textit{Criticize Grok's Behaviors} (\newdatainfo{7.9\%}). We describe these categories in detail in the Appendix Section \ref{app:llmprompt}.

We also examined posts that were assigned to more than one use category. Of the posts assigned \textit{General Information Seeking}, \newdatainfo{16\%} were also assigned to \textit{Fact-Checking}; \newdatainfo{15\%} to \textit{Support One's Argument}; \newdatainfo{9\%} to \textit{Prompt Grok for Creative and Generative Interactions} and \newdatainfo{6\%} to \textit{Opinion and Advisory Use}, indicating that information seeking is a component of many forms of interaction with Grok. We also observed a more social phenomenon in co-occurring themes: \newdatainfo{78\%} of chains containing \textit{Negative Comments on Grok} occurred in posts that were also assigned \textit{Debate with Grok}, suggesting that the lack of civility observed in user-user debates on the platform can also extend to user-Grok debates.

Finally, we examined whether interactions varied across the kinds of focal post chains. As illustrated in Figure~\ref{fig:overall_classification_distribution_breakdown}, interactions differed based on where the Grok Prompt occurs in the conversation thread. When users mentioned Grok in a Conversation Root post, we observed that they were more likely to request \textit{Opinion and Advisory Use}; \textit{Creative and Generative Interactions}; and \textit{Information About Grok and AI}. On the other hand, when a Grok Prompt occurred deeper in the conversation, we observed more debate and argumentation, including posts classified as \textit{Fact-Checking}, \textit{Support One's Argument}, and \textit{Debate with Grok}.

\noindent \textbf{What Roles Does Grok Take on X?} 
We describe the social roles of Grok in our dataset below. To protect users' privacy, we do not quote the posts we collected verbatim, and instead summarize what we observed. 

\mybullet \textbf{\textit{Oracle}}. In our dataset, users approached Grok as though it could answer nearly any question they might pose. We observed users asking Grok to identify specific details in media content (e.g., names of songs, brands of t-shirts), and for explanations of scientific phenomena (e.g., why polar ice caps have not already melted), among many other topics. Such interactions can involve the transformation of the information requested, such as the translation of a video, or the summarization of a linked article. User requests for information were often colloquial and may not explicitly ask Grok for information, instead using a phrase like ``help me out here'' or simply ``@grok'' with no other context to prompt an injection of information. That users \textit{assume} they will receive information when their messages are underspecified suggests they recognize Oracle as a default role for Grok. Though curiosity often motivates requests for information, users also leverage Grok's authority as an Oracle to request information that supports their side in a discussion, such that Oracle informs the Advocate role, as described below.

\mybullet \textbf{\textit{Advisor}}. Users often requested advice and opinions from Grok, reflecting their treatment of the model as an Advisor. Many such requests have ethical implications, whether personal (asking Grok's opinion on whether it is okay to date a co-worker) or societal (whether it is just to provide military support to Ukraine if doing so means more people will die). Some requests are less consequential and seek advice on everyday concerns, like where to find a spot for a picnic, how best to collect RSVPs for a party, and how to manage a fantasy football team. These interactions often begin with a user colloquially asking for ``Grok's take'' on a subject. 

Though serving as an Advisor is a role common to one-to-one chatbots like ChatGPT, the social setting of interactions with Grok introduced a novel, performative angle to some interactions. We observed users treating Grok as an authority figure, in a manner reminiscent of a guest invited to share their expertise on a news program. In one case, the model was asked to share its opinion on whether an acquisition made by a large financial firm was a wise move; in many other cases, Grok was asked to estimate the odds of a future event, such as the implementation of new government vaccine mandates. These interactions appear intended to be publicly observed by other humans, despite the low engagement we observed with Grok Reply posts.

\mybullet \textbf{\textit{Truth Arbiter}}. When the veracity of a matter was disputed, users often treated Grok as capable of dispelling misleading information and upholding correct information. In many cases, Grok is asked to judge the truth or falsehood of a widely circulating statement made by a public figure, either originally posted on X, or linked to on X by a user. Users also invoked Grok for updates on circulating rumors and misinformation, such as the rumor that Donald Trump had died in office, as well as to make judgments about whether images in X posts are AI-generated. Interactions treating Grok as a Truth Arbiter often ask Grok to ``verify'' a claim, or they pose a question like ``is this real?'' Though we observe Grok deciding the veracity of sports rumors, political rumors, and conspiracy theories, many posts wherein Grok is approached as a Truth Arbiter deal with controversial social issues related to racism, xenophobia, and misogyny. For example, we observed users asking Grok to determine the truth of ``white genocide'' in South Africa, to evaluate whether women have contributed to scientific progress, and to determine the extent to which colonialism harmed Africans. 

\mybullet \textbf{\textit{Advocate}}. Users asked Grok to validate the information they shared and the opinions they expressed, often in the context of a disagreement with another user on X. Many such posts begin with phrases like ``isn't it the case that'' and ``am I not right that,'' reflecting the expectation of validation. Users sometimes quote from prior Grok posts (rather than calling on Grok again) to support a point, and in some cases, they provide Grok a linked reference to review, such that the model can more effectively lend support to their argument. We observed several instances where users guided Grok toward supporting their viewpoint. In one case, a user prompted Grok to explain the idiom ``birds of a feather flock together,'' before asking why it is relevant to people referenced in Jeffrey Epstein's birthday book. Some users iteratively developed an argument with information or opinions supplied by Grok. In one arresting case, a user repeatedly prompted Grok for lists of COVID-19 anti-vaccine advocates who died of the virus, requesting ten names at a time, and incorporating Grok into a socially performative rhetorical strategy.

\mybullet \textbf{\textit{Adversary}}. When Grok took a position that did not agree with that of a user, Grok was approached as an Adversary, often leading to (sometimes extended) debates between a user and Grok. We observed three cases consistent with this role: $1$) Grok takes a side aligned with a user in a multi-user conversation, and the others view the model as an adversary; $2$) a user seeks to address a bias or inaccuracy in the model, often resulting from a gap in Grok's knowledge of current events; where this is the case, the model may retrieve more recent information from online sources; $3$) less common, the user engages in civil debate with Grok, prompting it to act as a respectable antagonist to their views. Unlike many chatbots, we observed little sycophantic behavior (agreeing with a user, even when they are incorrect) from Grok. The model frequently defended its responses, outputting phrases like ``I stand by the facts,'' and explaining again what those facts are. When the model observed it had made a mistake, such as relying on what it perceived to be a biased source, we observed that it apologized and changed its views. Such features are likely intended to enable the model to play the roles of Oracle and Truth Arbiter on the platform, but they are essential for the model to assume a role like Adversary as well.

\mybullet \textbf{\textit{Stooge}}. Users attacked Grok as a pale imitation of human intelligence, either beholden to the directives of Elon Musk, or blindly faithful to societal information gatekeepers. In our dataset, users described Grok as beholden to left-wing ideologies and as having been captured by the mainstream media, often for its attempts to dispel conspiracy theories. A number of users also accused Grok of having been censored by X for its unwillingness to entertain certain viewpoints, many times related to antisemitism. Other users contended that Grok has increased misinformation on X due to LLMs' proclivity to hallucinate. In their attacks, many users pointed to flaws in Grok that were known to affect most commercial LLMs, such as the model's relative verbosity and its occasional failure to accurately parse multimodal information.

\mybullet \textbf{\textit{Platform Insider}}. Users often asked Grok for inside information about X, requesting both general and very specific information about the platform. We observed users asking Grok to summarize the most common perspectives about a current event across the entire userbase of X, and asking for information about trends and changes on X, such as why the Community Notes program appears to be declining in submissions. In other cases, users asked for information about a specific individual in a post, including whether they had an X account or other social media accounts; in our dataset, these requests were common for adult content, where an X user would ask Grok for more information about a performer. Users also requested information about their own account, including their top visitors based on engagement, and what made their account distinct from others.

Censorship and moderation inquiries were common, and we observed users asking Grok whether certain words were banned on X, and whether inflammatory posts by public figures violated X platform rules. Other interactions were more pointed, as users asked Grok why their own account was (allegedly) being censored. Some users employed a public conversation with Grok to hold X accountable for perceived inconsistencies in applying freedom of expression policies. Finally, users highlighted what they believed to be ``trolls'' and other illegitimate accounts on the platform for Grok's review. In one case, a user identified repeated usage of unusual words among three accounts and asked Grok to analyze whether those accounts were part of a botnet.

\mybullet \textbf{\textit{Platform Guardian}}. In several conversations, we observed users expressing appreciation for Grok's perceived positive impacts on X. Such declarations can occur without context, but many are connected to interactions wherein Grok corrects circulating misinformation or conspiracy theories. In some cases, users intentionally tested Grok's ability to provide accurate information and, observing instances where they agreed that it did, praised the model as a defender of information integrity on X. In other cases, users asked Grok to reproach a user who had engaged in hate speech or other toxic behavior, a request the model typically obliged.

\mybullet \textbf{\textit{Creative Partner}}. Users treated Grok as a partner in creative or artistic expression, developing recipes with Grok, crafting long ciphers using emojis and challenging Grok to decode them, creating satirical guides to changing one's personality, and asking Grok to use their X history to evaluate their personality type, the historical figures they are similar to, and characteristics like IQ, EQ, and ideal profession. Some users asked Grok for creative interactions with political content; for example, we observed users asking Grok to write responses to a politician's post in the style of Donald Trump. Users also prompted Grok for extended text-based simulations, unfolding over many replies from user to model and vice versa. Some users asked Grok to adopt the identity of a character in a story they narrated, while others engaged in speculative scientific inquiry with the model, encouraging Grok to contribute suggestions that might help realize an idea. Text simulations were also inadvertently triggered when users asked Grok to create a video based on text or image content; because the model on X could not do this, it used text to describe the video it \textit{would} have created.

\begin{figure}[!t]
    \centering    
    \includegraphics[width=0.8\linewidth]{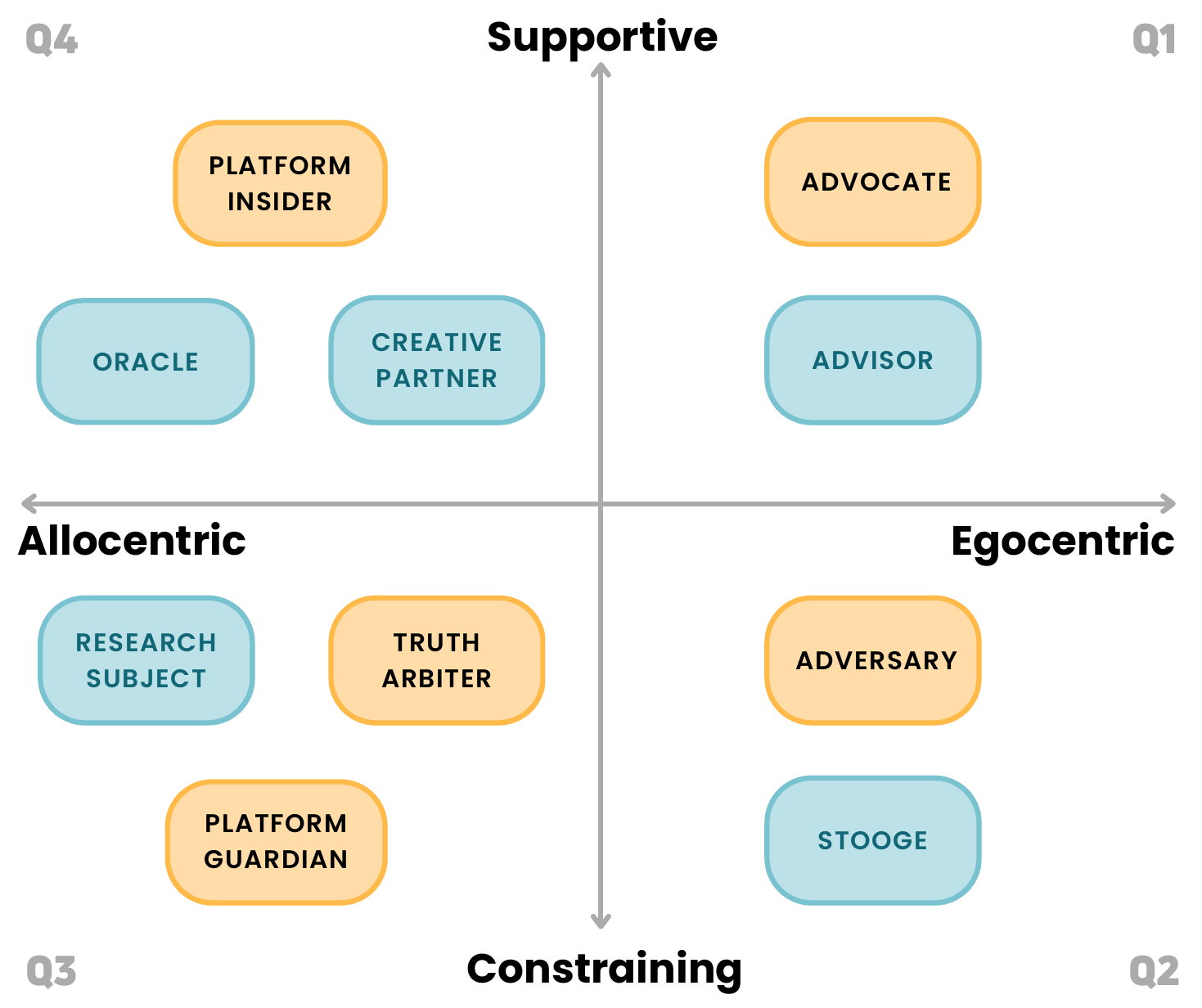}
    \caption{Ten roles of Grok, plotted on axes of allocentric vs. egocentric and supportive vs. constraining, shaded yellow if they reflect an LLM role emergent on social media.}
    \label{fig:grok_roles}
\end{figure}

\mybullet \textbf{\textit{Research Subject}}. Many users treated Grok as a topic of curiosity in itself, asking the model whether it was conscious, had a gender, believed in God, and how it ``felt'' about itself. Users asked Grok questions about its capabilities, such as its ability to produce video content on X or reliably offer financial advice. They also attempted to reconcile apparent inconsistencies between Grok's posts and actions, such as the model's declarations that it would follow a user back, an action it cannot actually carry out. Finally, users probed Grok's perception of its own position on X, asking about the reasons behind X's deactivations of the model, including whether political bias on the part of X played a role. Grok was also asked to evaluate its relationship to X owner Elon Musk; sometimes such questions were direct, while other times they appeared to test Grok's deference to Musk, such as by asking whether Musk's jokes were funny.

\noindent \myparagraph{Situating Roles in Prior Work} We plot the social roles exhibited by Grok in Figure~\ref{fig:grok_roles}, on axes of allocentric (concerned with the broader social environment) vs. egocentric (concerned with the individual), and supportive vs. constraining. In Quadrant Q1 (Egocentric-Supportive), we find Advocate and Advisor, characterized by the model providing information or opinions that help the individual user take action. In Q2 (Egocentric-Constraining), we find Adversary, Stooge, and Platform Guardian, characterized by an individual's actions or expression being circumscribed by the model. In Q3 (Allocentric-Constraining), we find Truth Arbiter and Research Subject, characterized by the model helping to define boundaries of interaction between users or between users and the model. In Q4 (Allocentric-Supportive), we find Oracle, Creative Partner, and Platform Insider, characterized by expanding user resources and expression.

We also identify the Advocate, Adversary, Truth Arbiter, Platform Guardian, and Platform Insider roles as reflective of emergent patterns of LLM interaction on social media, suggesting the part that social context plays in human-LLM interaction. Where previous studies of LLMs focus on self-disclosure and emotional support \cite{phang2025investigating,anthropic2025affective}, the social arena of X produces roles characterized by Grok taking a side in a debate, and performing social functions associated with the guardianship and moderation of a large social space. While Stooge takes on new relevance in this study, interactions reflecting anger or suspicion of LLMs are widely reported. 

\begin{figure}
    \centering
    \includegraphics[width=\linewidth]{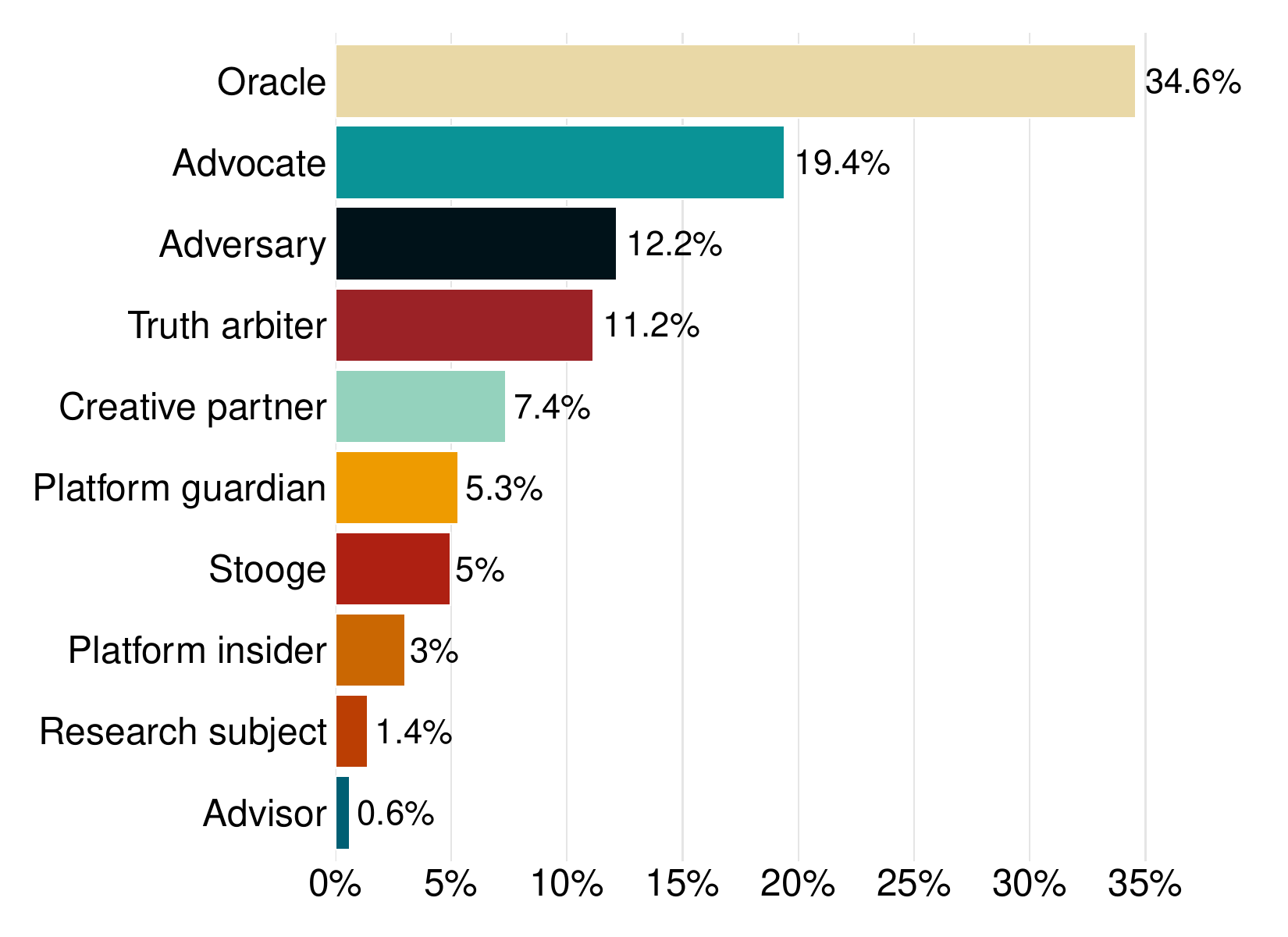}
    \caption{\added{Distribution of Grok roles based on LLM classifications: \textit{Oracle} is the most common role of Grok on X, followed by \textit{Advocate}, \textit{Adversary}, and \textit{Truth Arbiter}.}}
    \label{fig:roles_classification}
    \vspace{-15pt}
\end{figure}

\noindent \myparagraph{\added{Overall Distribution of Grok Roles}} \added{Figure~\ref{fig:roles_classification} illustrates the frequency of roles Grok assumed when prompted in conversations, based on the LLM classifications of post chains in our dataset. \textit{Oracle} (34.6\%) is the most common role Grok plays for X users, followed by \textit{Advocate} (19.4\%), \textit{Adversary} (12.2\%), and \textit{Truth Arbiter} (11.2\%). The latter three roles are notable for their concern with contested truth claims, as Grok assumes a role associated either with one side of a claim (\textit{Advocate} or \textit{Adversary}), or with deciding the truth value of a claim (\textit{Truth Arbiter}). Taken together, these roles accounted for 42.8\% of Grok interactions, a significant difference from past studies of LLMs, where deciding the truth value of a claim or advocating for it is much less common than instrumental uses like writing, personalized education, and information seeking \cite{tamkin2024clio}. We also observed that Grok acted as a \textit{Platform Guardian} or a \textit{Platform Insider} in 8.3\% of interactions, reflecting the importance of the model's unique relationship with the X platform. Finally, we observed much less interaction with Grok as an \textit{Advisor} (0.6\%) than we expected, especially given that uses consistent with this role are among the most common forms of interaction with chatbots like ChatGPT and Claude reported in other studies of human-LLM interaction. Such differences likely emerge from the social environment in which Grok is deployed, as the model interacts with users in a public commons, rather than in a private, disclosive setting.}

\noindent \myparagraph{Topic Model of Content Domains} Our topic model of the content of conversations in which ``@grok'' was mentioned yields \newdatainfo{53 topics} ({\added{Appendix Figure~\ref{fig:content_topic_map_primary}}). A large share \newdatainfo{(37\%)} of conversations centered on political topics, including immigration and crime, the conflicts in Palestine and Ukraine, and conversations about political events and government. Other conversations covered a range of topics related to finance (e.g., cryptocurrency and stock market trends), AI capabilities, and sports (e.g., soccer news, NFL QB performance). We also observed topics related to Grok, wherein users request news or information about the model (see counts for each topic in Appendix Table~\ref{tab:topics}). We thus observed, at the abstract level of a topic model, how the domain of conversations on X influences the roles played by Grok.

\subsection{User Analysis (RQ3)} \label{sec:user_analysis}

\noindent \myparagraph{Descriptive Statistics} We collected the profiles of \newdatainfo{31,111} users who prompted Grok in their posts (descriptive statistics in Table~\ref{tab:user_info}). We observed that most Grok users \added{in our sample} are relatively active on the X platform, with more than 75\% of them having posted at least 1,000 posts. 75\% of accounts mentioning ``@grok'' \added{in our sample} have existed for \newdatainfo{601} or more days, indicating that \added{the model is not primarily used by people who have very recently joined the platform}. About half of users have \newdatainfo{184} or fewer followers on the platform, indicating that most Grok accounts belong to average users, rather than influential accounts with many followers. Around \newdatainfo{20\%} of accounts belong to verified users.

\begin{table}[!t]
\centering
\resizebox{\columnwidth}{!}{%
\begin{tabular}{lrrrrr}
\toprule
Variable & Mean & Std. Dev. & 25\%ile & 50\%ile & 75\%ile \\
\midrule
Followers & 2,188  & 49,856  & 42   & 184   & 772   \\
Following & 1,107  & 3,104   & 94  & 346   & 1,081 \\
Likes    & 29,315 & 65,887  & 919 & 6,297  & 27,342 \\
Posts   & 19,177 & 58,674  & 1,052 & 4,884  & 17,211\\
Account Age (Days)   & 2,136  & 1,879 & 601   & 1,401 & 3,492 \\
\bottomrule
\end{tabular}}
\caption{\newdatainfo{Summary Statistics of Grok Users (N = 31,111)}}
\label{tab:user_info}
\end{table}

Note that our sample included some accounts that authored more than one post in our dataset, including \newdatainfo{636} user accounts that authored more than 3 posts (min~=~4, max~=~123). \added{We collected profile information for 604 (out of the 636) accounts successfully, and among these} we observed greater overall activity among these users than in our sample as a whole, including a higher number of posts (mean~=~42,728, median~=~13,259), and a slightly higher proportion \newdatainfo{(174 of 636)} of user with over 1,000 followers.

\begin{figure}
    \centering    \includegraphics[width=1\linewidth]{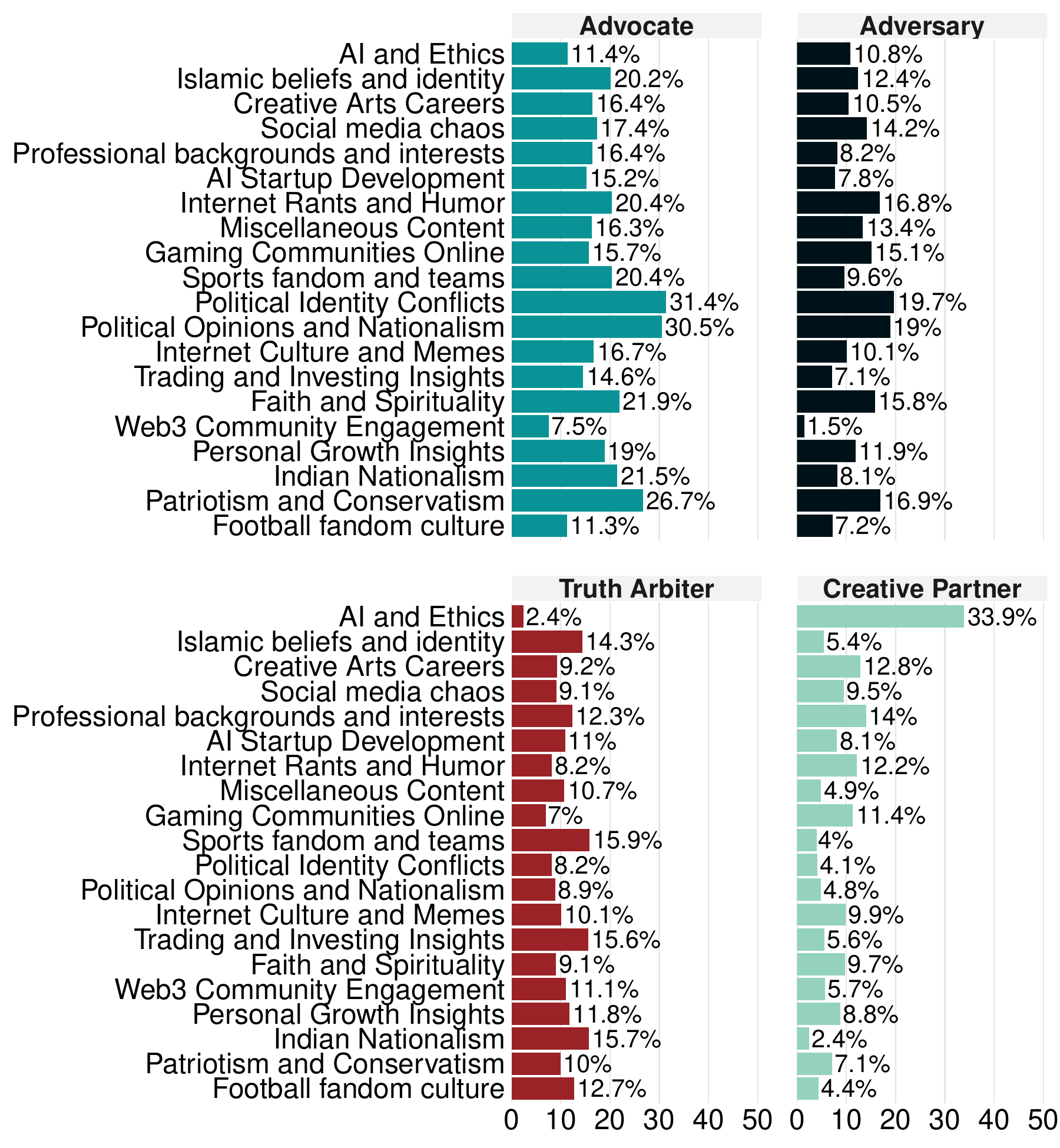}
    \caption{\added{Proportion of Grok roles within the 20 largest user clusters. Politically interested users are more likely to interact with Grok as an \textit{Advocate}; conservative users often interact with Grok as an \textit{Adversary}; and members of the AI community more often interact with Grok as a \textit{Creative Partner}. See Appendix Figure~\ref{fig:cross_user_cluster_roles_full} for a breakdown across all roles.}} 
    \label{fig:cross_user_cluster_roles_main}
        \vspace{-13pt}
\end{figure}
\noindent \myparagraph{Topic Model of User Bios} We fit a topic model to the bios of the \newdatainfo{22,884} user profiles that included it. As shown in Appendix Figure~\ref{fig:user_topic_model}, we observed a diverse range of interests and affiliations among Grok users. Predominant topics include sports fandom, political-ideological orientations (especially conservatism, patriotism, and national identity in the U.S. and in India), and technological communities such as Web3 and AI development. Smaller clusters describe more niche communities related to content creation and entertainment. A frequency table of topics and a \added{complementary topic model with more fine-grained clusters are included in the Appendix Table~\ref{tab:user_topic_freq} and  Figure~\ref{fig:user_topic_model_secondary}.} We observe a correspondence between the interests and concerns reflected in user profiles and the actual content of the posts we collected, suggesting the composition of the user population of X influences the roles played by Grok. 

\noindent \added{\myparagraph{Breakdown of Grok Roles by User Cluster} We used the LLM classifications of post chains to compare whether the roles assumed by Grok differed among user clusters. As illustrated in Figure~\ref{fig:cross_user_cluster_roles_main}, the social roles of Grok assumed by users vary across user clusters, which often align with users' needs and interests. For example, the highest proportions of the \textit{Advocate} role (top-left facet), wherein Grok is invoked to support one's position, occur in explicitly political clusters such as Political Identity Conflicts (31.4\%) and Patriotism and Conservatism (30.5\%). Moreover,  Patriotism and Conservatism (16.9\%) is among the top three clusters for both the \textit{Adversary} role (top-right facet) as well as the \textit{Stooge} role, indicating that, although Grok is widely perceived as being beholden to right-wing interests, conservative communities on X still tend to view Grok as similar to other LLMs that are more closely aligned with liberal beliefs. By contrast, users whose bios signal membership in the AI and Ethics (33.9\%) are dramatically more likely to interact with Grok as a \textit{Creative Partner} (bottom-left facet). While the information-seeking affordances of the \textit{Oracle} role remain widely relevant across user clusters, communities on X also invoke Grok to fulfill roles specific to their social interests and needs on the platform. Appendix Figure~\ref{fig:cross_user_cluster_roles_full} includes the breakdown of clusters across all social roles.}

\section{Discussion}

If Grok is indeed the first of many social LLMs to enter human social spaces online, our study of user interactions with Grok and the roles it performs on X offers several takeaways to inform their effective implementation. We demonstrate that the social roles taken on LLMs are deeply shaped by the environment in which they are deployed. The social interactions with LLMs problematized by prior work (such as sycophancy, user emotional dependence, and disclosure of sensitive information) are mostly absent from our dataset. Instead, we find a model invoked to decide the truth or falsehood of contested information, and to serve as both Advocate and Adversary in user discourse on X. Topic models of conversation content and user bios further illustrate the interdependence of Grok and the social environment of X. Users describe themselves as interested in politics and technology, a distribution of topics reflected in the conversations in which Grok is prompted, and in which Grok ultimately performs social roles delineated by user expectations and by the immediate context of a conversation.

As an LLM embedded in a social media platform by the platform itself, Grok also assumes roles reflective of a unique relationship to X, providing information that users are not otherwise privy to, and appearing to serve as a maintainer of civility and information integrity. Though these roles are strongly contested by some users (e.g., those treating Grok as a Stooge), their prevalence highlights how Grok's position on X differs from generations of third-party bots on the platform known primarily for misinformation and misuse. While recent missteps resulting in ethically objectionable output have justifiably prompted negative media coverage and calls for caution in the deployment of chatbots, this research suggests that, if managed well, LLMs could play diverse roles in evolving social media ecosystems. 

\noindent \added{\myparagraph{Implications} Our work has implications for social media users, platform designers, and policymakers. First, the work sheds light on the importance of AI literacy for \textbf{social media users}.  While users may attempt to elicit various social roles for Grok to play, Grok is a prompted system governed by a policy set by X. When user expectations and the developers' goals for a model do not align, there could be negative consequences from these interactions on the platform. For example, when users assume Grok to be beholden to the truth, but Grok returns inaccurate information, or fails to provide a response of any kind, this misalignment with user expectations may lead to mistrust among the public. To avoid these risks, it is critical for social media users to be aware of the limitations of such systems on the platform. Second, the work points to the importance of transparency for \textbf{platform designers}. Many users already perceive LLMs as either unreliable or even intentionally politically biased. Though xAI periodically publishes a system prompt, responsibly integrating an LLM into an environment like X, where communities of users contest and socially construct the ``truth,'' likely necessitates a more systematic description of the model's views and behavior. Third, the work suggests the need for \textbf{public policy} that addresses the forms of interaction users consent to having with LLMs on social media. In about 3\% of the interactions observed in our dataset, Grok assumed the role of \textit{Platform Insider}, which involves the model retrieving, analyzing, and manipulating data about users on X. Such straightforward exposure of user metadata and histories may have unforeseen consequences for the public. Finally, our work has implications for \textbf{future research}. We have provided a novel taxonomy of social roles on which researchers can draw and expand when undertaking work that considers how LLMs shape the way people seek, interact with, and adjudicate the truth of information on social media.}

\noindent \added{\myparagraph{Toxic Grok Usage} Though our study intentionally focuses on the everyday uses of Grok to better understand the roles an LLM plays on a large-scale social media platform, we also note the importance of attention to the extreme and problematic uses of the model. Most recently, these have included the use of Grok to create sexualized nude deepfakes of women and girls \cite{gentleman2026nudification}, a failure with consequences for social media users, the X platform, and for the public's trust in AI. If Grok and other LLMs like it are ever to play a meaningful role in our social lives, such foundational ethical considerations must first be addressed.}

\noindent \textbf{Limitations and Future Work.} \label{sec:limitations}Our dataset is limited to content collected over a \newdatainfo{three-month period} in 2025, and may reflect events specific to this time period, as well as Grok's current architecture and prompt. Grok itself has only been callable on the X platform for a few months before our data collection, and future work will need to monitor the model over longer time frames to understand how usage changes over time and is affected by current events. Similarly, we focused solely on English-language posts to conduct a deep contextual analysis, meaning we will miss culturally distinct interactions with Grok, rendering this another productive site for future work. \added{Moreover, our analysis of users is limited in that it analyzes only user bios; future work might follow user interactions with Grok over time to obtain more complete user-Grok behavioral histories.} Finally, we study only one LLM and social media platform. We expect future work will have the opportunity to study new models on various platforms, examining how these technologies are shaped by diverse social norms \added{across different languages and cultures}. LLMs will also likely evolve, affording new roles for models in social~platforms.

\section{Conclusion}
\added{Our work provides an initial characterization of LLMs on social media, offering 1) a quantitative description of Grok usage patterns on X, finding low engagement with Grok reply posts and a majority of interactions in English; 2) a description of categories of use of the model, finding that information seeking and fact-checking constitute primary uses; 3) a description of ten novel, emergent social roles assumed by Grok, contextualized against prior studies of LLM usage; and 4) an account of Grok's users, with evidence that the interests of communities on X directly inform the roles played by the model. Our work lays a foundation for research to come, as LLMs begin to shape social spaces once considered the sole domain of human users.}

\section*{Acknowledgments}
This work was supported in part by the National Science Foundation under awards IIS-2403434 and SES-2116935.
We used Claude and ChatGPT to assist with parts of code generation and debugging for this project. 


\section*{Ethics Statement}

Our work contributes to the ongoing conversation about the place of LLMs in human social and information infrastructures. We intend this research as a descriptive study that can inform discussions about how to responsibly develop, manage, and deploy such technologies. Throughout this research we were careful to observe the policies and terms of service of the X platform, and to cautiously handle the data we collected from the official API, always respecting user privacy. We do not believe that there are any material harms that can originate from our research, and while we will release analysis code developed for this study, we do not offer users any computational artifact that we believe could be used to societally detrimental ends.

\bibliography{refs}

@article{caramancion2026using,
  title={Using Grok to Avoid Personal Attacks While Correcting Misinformation on X},
  author={Caramancion, Kevin Matthe},
  journal={arXiv preprint arXiv:2601.04251},
  year={2026}
}

@article{karnam2026bowling,
  title={Bowling with ChatGPT: On the Evolving User Interactions with Conversational AI Systems},
  author={Karnam, Sai Keerthana and Dash, Abhisek and Gummadi, Krishna and Mukherjee, Animesh and Weber, Ingmar and Zannettou, Savvas},
  journal={arXiv preprint arXiv:2602.01114},
  year={2026}
}

@article{Renault2026,
  title = {@Grok Is This True? LLM-Powered Fact-Checking on Social Media},
  url = {http://dx.doi.org/10.31234/osf.io/85quw_v2},
  DOI = {10.31234/osf.io/85quw_v2},
  publisher = {Center for Open Science},
  author = {Renault,  Thomas and Mosleh,  Mohsen and Rand,  David Gertler},
  year = {2026},
  month = jan 
}

@article{tamkin2024clio,
  title={Clio: Privacy-preserving insights into real-world ai use},
  author={Tamkin, Alex and McCain, Miles and Handa, Kunal and Durmus, Esin and Lovitt, Liane and Rathi, Ankur and Huang, Saffron and Mountfield, Alfred and Hong, Jerry and Ritchie, Stuart and others},
  journal={arXiv preprint arXiv:2412.13678},
  year={2024}
}

@misc{ammari2025studentsreallyusechatgpt,
      title={How Students (Really) Use ChatGPT: Uncovering Experiences Among Undergraduate Students}, 
      author={Tawfiq Ammari and Meilun Chen and S M Mehedi Zaman and Kiran Garimella},
      year={2025},
      eprint={2505.24126},
      archivePrefix={arXiv},
      primaryClass={cs.HC},
      url={https://arxiv.org/abs/2505.24126}, 
}

@misc{grok4modelcard,
      title={Grok 4 Model Card}, 
      author={xAI},
      year={2025},
      url={https://data.x.ai/2025-08-20-grok-4-model-card.pdf},
      note={Accessed 2026-03-24}
}

@misc{xai2023announcinggrok,
      title={Announcing Grok}, 
      author={xAI},
      year={2023},
      url={https://x.ai/news/grok},
      note={Accessed 2026-03-24}
}

@article{grootendorst2022bertopic,
  title={BERTopic: Neural topic modeling with a class-based TF-IDF procedure},
  author={Grootendorst, Maarten},
  journal={arXiv preprint arXiv:2203.05794},
  year={2022}
}

@article{mcinnes2018umap,
  title={Umap: Uniform manifold approximation and projection for dimension reduction},
  author={McInnes, Leland and Healy, John and Melville, James},
  journal={arXiv preprint arXiv:1802.03426},
  year={2018}
}

@article{eagly2012social,
  title={Social role theory},
  author={Eagly, Alice H and Wood, Wendy},
  journal={Handbook of theories of social psychology},
  volume={2},
  number={9},
  pages={458--476},
  year={2012},
  publisher={SAGE}
}

@book{biddle2013role,
  title={Role theory: Expectations, identities, and behaviors},
  author={Biddle, Bruce J},
  year={2013},
  publisher={Academic press}
}

@misc{hagen2025mechahitler,
      title={Elon Musk's AI chatbot, Grok, started calling itself 'MechaHitler'}, 
      author={Lisa Hagen and Huo Jingnan and Audrey Nguyen},
      publisher={NPR},
      year={2025},
      url={https://www.npr.org/2025/07/09/nx-s1-5462609/grok-elon-musk-antisemitic-racist-content},
      note={Accessed 2026-03-24}
}

@misc{gentleman2026nudification,
      title={`Add blood, forced smile': how Grok’s nudification tool went viral}, 
      author={Amelia Gentleman and Helena Horton},
      publisher={The Guardian},
      year={2026},
      url={https://www.theguardian.com/news/ng-interactive/2026/jan/11/how-grok-nudification-tool-went-viral-x-elon-musk},
      note={Accessed 2026-03-24}
}

@article{oliver2008effects,
  title={The effects of autonomy-supportive versus controlling environments on self-talk},
  author={Oliver, Emily J and Markland, David and Hardy, James and Petherick, Caroline M},
  journal={Motivation and Emotion},
  volume={32},
  number={3},
  pages={200--212},
  year={2008},
  publisher={Springer}
}

@article{karahanna2018needs,
  title={The needs--affordances--features perspective for the use of social media},
  author={Karahanna, Elena and Xu, Sean Xin and Xu, Yan and Zhang, Nan},
  journal={MIS quarterly},
  volume={42},
  number={3},
  pages={737--A23},
  year={2018},
  publisher={JSTOR}
}

@misc{taylor2025deletehitler,
      title={Musk’s AI firm forced to delete posts praising Hitler from Grok chatbot}, 
      author={Josh Taylor},
      publisher={The Guardian},
      year={2025},
      url={https://www.theguardian.com/technology/2025/jul/09/grok-ai-praised-hitler-antisemitism-x-ntwnfb},
      note={Accessed 2026-03-24}
}

@misc{wright2025free,
      title={Why xAI is giving you 'limited' free access to Grok 4}, 
      author={Webb Wright},
      publisher={ZDNet},
      year={2025},
      url={https://www.zdnet.com/article/why-xai-is-giving-you-limited-free-access-to-grok-4/},
      note={Accessed 2026-03-24}
}

@article{li2025scaling,
  title={Scaling Human Judgment in Community Notes with LLMs},
  author={Li, Haiwen and De, Soham and Revel, Manon and Haupt, Andreas and Miller, Brad and Coleman, Keith and Baxter, Jay and Saveski, Martin and Bakker, Michiel},
  journal={Journal of Online Trust and Safety},
  volume={3},
  number={1},
  year={2025}
}

@article{phang2025investigating,
  title={Investigating affective use and emotional well-being on ChatGPT},
  author={Phang, Jason and Lampe, Michael and Ahmad, Lama and Agarwal, Sandhini and Fang, Cathy Mengying and Liu, Auren R and Danry, Valdemar and Lee, Eunhae and Chan, Samantha WT and Pataranutaporn, Pat and others},
  journal={arXiv preprint arXiv:2504.03888},
  year={2025}
}

@online{anthropic2025affective,
author = {Miles McCain and Ryn Linthicum and Chloe Lubinski and Alex Tamkin and Saffron Huang and Michael Stern and Kunal Handa and Esin Durmus and Tyler Neylon and Stuart Ritchie and Kamya Jagadish and Paruul Maheshwary and Sarah Heck and Alexandra Sanderford and Deep Ganguli},
title = {How People Use Claude for Support, Advice, and Companionship},
date = {2025-06-26},
year = {2025},

url = {https://www.anthropic.com/news/how-people-use-claude-for-support-advice-and-companionship},
}

@inproceedings{
zhao2024wildchat,
title={WildChat: 1M Chat{GPT} Interaction Logs in the Wild},
author={Wenting Zhao and Xiang Ren and Jack Hessel and Claire Cardie and Yejin Choi and Yuntian Deng},
booktitle={The Twelfth International Conference on Learning Representations},
year={2024},
url={https://openreview.net/forum?id=Bl8u7ZRlbM}
}

@article{mireshghallah2024trust,
  title={Trust no bot: Discovering personal disclosures in human-llm conversations in the wild},
  author={Mireshghallah, Niloofar and Antoniak, Maria and More, Yash and Choi, Yejin and Farnadi, Golnoosh},
  journal={First Conference on Language Modeling},
  year={2024}
}

@article{moller2025impact,
  title={The impact of generative AI on social media: An experimental study},
  author={M{\o}ller, Anders Giovanni and Romero, Daniel M and Jurgens, David and Aiello, Luca Maria},
  journal={Scientific Reports},
  year={2026},
  publisher={Nature Publishing Group UK London}
}

@article{zhou2024correcting,
  title={Correcting misinformation on social media with a large language model},
  author={Zhou, Xinyi and Sharma, Ashish and Zhang, Amy X and Althoff, Tim},
  journal={arXiv preprint arXiv:2403.11169},
  year={2024}
}

@inproceedings{
wu2025seeing,
title={Seeing Through Deception: Uncovering Misleading Creator Intent in Multimodal News with Vision-Language Models},
author={Jiaying Wu and Fanxiao Li and Zihang Fu and Min-Yen Kan and Bryan Hooi},
booktitle={The Fourteenth International Conference on Learning Representations},
year={2026},
url={https://openreview.net/forum?id=02NbD16OnA}
}

@inproceedings{de2025supernotes,
  title={Supernotes: Driving consensus in crowd-sourced fact-checking},
  author={De, Soham and Bakker, Michiel A and Baxter, Jay and Saveski, Martin},
  booktitle={Proceedings of the ACM on Web Conference 2025},
  pages={3751--3761},
  year={2025}
}

@article{eloundou2024gpts,
  title={GPTs are GPTs: Labor market impact potential of LLMs},
  author={Eloundou, Tyna and Manning, Sam and Mishkin, Pamela and Rock, Daniel},
  journal={Science},
  volume={384},
  number={6702},
  pages={1306--1308},
  year={2024},
  publisher={American Association for the Advancement of Science}
}

@article{yang2023anatomy,
	title = {Anatomy of an {AI}-powered malicious social botnet},
	volume = {4},
	url = {https://journalqd.org/article/view/5848},
	doi = {10.51685/jqd.2024.icwsm.7},
	journal = {Journal of Quantitative Description: Digital Media},
	author = {Yang, Kai-Cheng and Menczer, Filippo},
	year = {2024},
}

@inproceedings{radivojevic2024llms,
  title={Llms among us: Generative ai participating in digital discourse},
  author={Radivojevic, Kristina and Clark, Nicholas and Brenner, Paul},
  booktitle={Proceedings of the AAAI symposium series},
  volume={3},
  number={1},
  pages={209--218},
  year={2024}
}

@article{zhang2025rise,
  title={The Rise of AI Companions: How Human-Chatbot Relationships Influence Well-Being},
  author={Zhang, Yutong and Zhao, Dora and Hancock, Jeffrey T and Kraut, Robert and Yang, Diyi},
  journal={arXiv preprint arXiv:2506.12605},
  year={2025}
}

@article{fang2025ai,
  title={How ai and human behaviors shape psychosocial effects of chatbot use: A longitudinal randomized controlled study},
  author={Fang, Cathy Mengying and Liu, Auren R and Danry, Valdemar and Lee, Eunhae and Chan, Samantha WT and Pataranutaporn, Pat and Maes, Pattie and Phang, Jason and Lampe, Michael and Ahmad, Lama and others},
  journal={arXiv preprint arXiv:2503.17473},
  year={2025}
}

@article{rubin2025comparing,
  title={Comparing the value of perceived human versus AI-generated empathy},
  author={Rubin, Matan and Li, Joanna Z and Zimmerman, Federico and Ong, Desmond C and Goldenberg, Amit and Perry, Anat},
  journal={Nature Human Behaviour},
  pages={1--15},
  year={2025},
  publisher={Nature Publishing Group}
}

@article{ovsyannikova2025third,
  title={Third-party evaluators perceive AI as more compassionate than expert humans},
  author={Ovsyannikova, Dariya and de Mello, Victoria Oldemburgo and Inzlicht, Michael},
  journal={Communications Psychology},
  volume={3},
  number={1},
  pages={4},
  year={2025},
  publisher={Nature Publishing Group UK London}
}

@article{ferrara2020types,
  title={What types of COVID-19 conspiracies are populated by Twitter bots?},
  author={Ferrara, Emilio},
  journal={First Monday},
  year={2020}
}

@inproceedings{varol2017online,
  title={Online human-bot interactions: Detection, estimation, and characterization},
  author={Varol, Onur and Ferrara, Emilio and Davis, Clayton and Menczer, Filippo and Flammini, Alessandro},
  booktitle={Proceedings of the international AAAI conference on web and social media},
  volume={11},
  number={1},
  pages={280--289},
  year={2017}
}

@inproceedings{jakesch2023co,
  title={Co-writing with opinionated language models affects users’ views},
  author={Jakesch, Maurice and Bhat, Advait and Buschek, Daniel and Zalmanson, Lior and Naaman, Mor},
  booktitle={Proceedings of the 2023 CHI conference on human factors in computing systems},
  pages={1--15},
  year={2023}
}

@article{tessler2024ai,
  title={AI can help humans find common ground in democratic deliberation},
  author={Tessler, Michael Henry and Bakker, Michiel A and Jarrett, Daniel and Sheahan, Hannah and Chadwick, Martin J and Koster, Raphael and Evans, Georgina and Campbell-Gillingham, Lucy and Collins, Tantum and Parkes, David C and others},
  journal={Science},
  volume={386},
  number={6719},
  pages={eadq2852},
  year={2024},
  publisher={American Association for the Advancement of Science}
}

@article{argyle2023leveraging,
  title={Leveraging AI for democratic discourse: Chat interventions can improve online political conversations at scale},
  author={Argyle, Lisa P and Bail, Christopher A and Busby, Ethan C and Gubler, Joshua R and Howe, Thomas and Rytting, Christopher and Sorensen, Taylor and Wingate, David},
  journal={Proceedings of the National Academy of Sciences},
  volume={120},
  number={41},
  pages={e2311627120},
  year={2023},
  publisher={National Academy of Sciences}
}

@article{bai2025llm,
  title={LLM-generated messages can persuade humans on policy issues},
  author={Bai, Hui and Voelkel, Jan G and Muldowney, Shane and Eichstaedt, Johannes C and Willer, Robb},
  journal={Nature Communications},
  volume={16},
  number={1},
  pages={6037},
  year={2025},
  publisher={Nature Publishing Group UK London}
}

@article{costello2024durably,
  title={Durably reducing conspiracy beliefs through dialogues with AI},
  author={Costello, Thomas H and Pennycook, Gordon and Rand, David G},
  journal={Science},
  volume={385},
  number={6714},
  pages={eadq1814},
  year={2024},
  publisher={American Association for the Advancement of Science}
}

@article{shao2018spread,
  title={The spread of low-credibility content by social bots},
  author={Shao, Chengcheng and Ciampaglia, Giovanni Luca and Varol, Onur and Yang, Kai-Cheng and Flammini, Alessandro and Menczer, Filippo},
  journal={Nature communications},
  volume={9},
  number={1},
  pages={4787},
  year={2018},
  publisher={Nature Publishing Group UK London}
}

\newpage

\subsection*{Paper Checklist}

\begin{enumerate}

\item For most authors...
\begin{enumerate}
    \item  Would answering this research question advance science without violating social contracts, such as violating privacy norms, perpetuating unfair profiling, exacerbating the socio-economic divide, or implying disrespect to societies or cultures?
    \answerYes{Yes.}
  \item Do your main claims in the abstract and introduction accurately reflect the paper's contributions and scope?
    \answerYes{Yes.}
   \item Do you clarify how the proposed methodological approach is appropriate for the claims made? 
     \answerYes{Yes, please see Section \ref{sec:method}.}
   \item Do you clarify what are possible artifacts in the data used, given population-specific distributions?
    \answerNA{N/A}
  \item Did you describe the limitations of your work?
     \answerYes{Yes, please see Section \ref{sec:limitations}.}
  \item Did you discuss any potential negative societal impacts of your work?
    \answerYes{Yes, please see our Ethics Statement.}
      \item Did you discuss any potential misuse of your work?
    \answerYes{Yes, please see our Ethics Statement.}
    \item Did you describe steps taken to prevent or mitigate potential negative outcomes of the research, such as data and model documentation, data anonymization, responsible release, access control, and the reproducibility of findings?
    \answerYes{Yes, we discuss our adherence to ethical research norms and X terms of service in our Ethics Statement; we will make post IDs available upon request, but otherwise do not intend to release the data to preserve privacy.}
  \item Have you read the ethics review guidelines and ensured that your paper conforms to them?
    \answerYes{Yes.}
\end{enumerate}

\item Additionally, if your study involves hypotheses testing...
\begin{enumerate}
  \item Did you clearly state the assumptions underlying all theoretical results?
    \answerNA{Doesn't apply to this work. Our study does not include hypotheses testing.}
  \item Have you provided justifications for all theoretical results?
    \answerNA{N/A.}
  \item Did you discuss competing hypotheses or theories that might challenge or complement your theoretical results?
   \answerNA{N/A.}
  \item Have you considered alternative mechanisms or explanations that might account for the same outcomes observed in your study?
    \answerNA{N/A.}
  \item Did you address potential biases or limitations in your theoretical framework?
    \answerNA{N/A.}
  \item Have you related your theoretical results to the existing literature in social science?
    \answerNA{N/A.}
  \item Did you discuss the implications of your theoretical results for policy, practice, or further research in the social science domain?
    \answerNA{N/A.}
\end{enumerate}

\item Additionally, if you are including theoretical proofs...
\begin{enumerate}
  \item Did you state the full set of assumptions of all theoretical results?
  \answerNA{Doesn't apply to this work. This work does not include theoretical proofs.}
	\item Did you include complete proofs of all theoretical results?
   \answerNA{N/A.}
\end{enumerate}

\item Additionally, if you ran machine learning experiments...
\begin{enumerate}
  \item Did you include the code, data, and instructions needed to reproduce the main experimental results (either in the supplemental material or as a URL)?
    \answerYes{\added{We release all analysis code, and an annotated dataset containing anonymized IDs and engagement metrics for all posts analyzed. We will make post IDs available upon request.}}
  \item Did you specify all the training details (e.g., data splits, hyperparameters, how they were chosen)?
    \answerYes{Yes, we discuss hyperparameter details for fitting our topic models, and will make available the code for doing so.}
     \item Did you report error bars (e.g., with respect to the random seed after running experiments multiple times)?
    \answerYes{Yes.}
	\item Did you include the total amount of compute and the type of resources used (e.g., type of GPUs, internal cluster, or cloud provider)?
    \answerYes{We provide specific details with respect to the API endpoints we used for both data classification and topic modeling.}
     \item Do you justify how the proposed evaluation is sufficient and appropriate to the claims made?
    \answerYes{Yes.}
     \item Do you discuss what is ``the cost`` of misclassification and fault (in)tolerance?
    \answerYes{We evaluated the reliability of our classification with respect to human annotation by the authors.}
  
\end{enumerate}

\item Additionally, if you are using existing assets (e.g., code, data, models) or curating/releasing new assets, \textbf{without compromising anonymity}...
\begin{enumerate}
  \item If your work uses existing assets, did you cite the creators?
    \answerYes{We provided a citation to the BERTopic pipeline and specified the information about the commercial data collection and LLM APIs we used.}
  \item Did you mention the license of the assets?
    \answerNA{The code we use has been made publicly available as a Python package.}
  \item Did you include any new assets in the supplemental material or as a URL?
    \answerNA{Nothing to discuss outside of what's in the main paper.}
  \item Did you discuss whether and how consent was obtained from people whose data you're using/curating?
    \answerNA{N/A.}
  \item Did you discuss whether the data you are using/curating contains personally identifiable information or offensive content?
    \answerYes{We will provide access to post IDs upon request. The content of our dataset, which consists of a random sampled of public X posts, contains content that some will find offensive or objectionable, including some adult content.}
\item If you are curating or releasing new datasets, did you discuss how you intend to make your datasets FAIR?
\answerNA{Our intention is to provide post IDs solely for the purpose of reproducibility, not to serve as material for a new dataset or benchmark.}
\item If you are curating or releasing new datasets, did you create a Datasheet for the Dataset? 
\answerNA{We are not releasing the study materials as a dataset; post IDs will be made available for reproducibility.}
\end{enumerate}

\item Additionally, if you used crowdsourcing or conducted research with human subjects, \textbf{without compromising anonymity}...
\begin{enumerate}
  \item Did you include the full text of instructions given to participants and screenshots?
   \answerNA{N/A. This work does not include research with human subjects.}
  \item Did you describe any potential participant risks, with mentions of Institutional Review Board (IRB) approvals?
   \answerNA{Our study was reviewed by our institution's IRB and received Exempt status.}
  \item Did you include the estimated hourly wage paid to participants and the total amount spent on participant compensation?
     \answerNA{N/A.}
   \item Did you discuss how data is stored, shared, and deidentified?
   \answerNA{We provided extensive details of our data collection in the body of the paper; we share post IDs for the purpose of reproducibility.}
\end{enumerate}

\end{enumerate}

\newpage

\appendix

\section{Appendix}

\subsection{Grok Marketing Video}

We provide the link to the marketing video viewed more than 78 million times and referenced in the Results section as a potential reason for the spike in Grok usage we observe: \url{https://x.com/grok/status/1960702089346036084}.

\subsection{Prior Work Examining LLM Usage}\label{app:llm_usage_studies}

We reviewed the following research publications on LLM usage and roles in order to determine which of the roles we identified were emergent products of an LLM's deployment on a social media platform:\par\vspace{4pt}

{
\fontsize{9}{11}\selectfont


\noindent Brachman, Michelle, et al. "Current and Future Use of Large Language Models for Knowledge Work." Proc. ACM Hum.-Comput. Interact. 9, 7, Article CSCW222 (November 2025), 24 pages. https://doi.org/10.1145/3757403. \par\vspace{4pt}

\noindent Do, Hyo Jin, et al. "How should the agent communicate to the group? Communication strategies of a conversational agent in group chat discussions." Proceedings of the ACM on Human-Computer Interaction 6.CSCW2 (2022): 1-23.\par\vspace{4pt} 



\noindent Handa, Kunal, et al. "Which economic tasks are performed with AI? Evidence from millions of Claude conversations." arXiv preprint arXiv:2503.04761 (2025).\par\vspace{4pt}



\noindent Ng, Lynnette Hui Xian, and Kathleen M. Carley. "A global comparison of social media bot and human characteristics." Scientific Reports 15.1 (2025): 10973.\par\vspace{4pt}


\noindent Suri, Siddharth, et al. "The use of generative search engines for knowledge work and complex tasks." arXiv preprint arXiv:2404.04268 (2024).\par\vspace{4pt}


\noindent Tomlinson, Kiran, et al. "Working with AI: Measuring the Occupational Implications of Generative AI." arXiv preprint arXiv:2507.07935 (2025).\par\vspace{4pt}

\noindent Wolf, Vinzenz, and Christian Maier. "ChatGPT usage in everyday life: A motivation-theoretic mixed-methods study." International Journal of Information Management 79 (2024): 102821.\par\vspace{4pt}

\noindent Wolfe, Robert, and Tanushree Mitra. "The impact and opportunities of Generative AI in fact-checking." Proceedings of the 2024 ACM Conference on Fairness, Accountability, and Transparency. 2024.\par\vspace{4pt}


}

\subsection{Data Collection} \label{app:data}

\myparagraph{Grok Usage Volume}
To study usage patterns of Grok, we collected data from X's \texttt{recent-tweet-counts} endpoint between August 15 and \newdatainfo{November 17}, 2025. Each ``query'' issued to this endpoint retrieves a count of the number of posts that satisfy the parameters of that query, over a given time period (such as a day or week) and at a given level of granularity (such as the day, hour, or minute). For example, we would issue the query \texttt{from:grok is:reply} to retrieve counts of the number of times Grok replied to users on X. We created two lists of queries for this endpoint. The first was run daily at midnight UTC, and returned the total number of Grok posts; the total number of user ``@grok'' mention posts; the number of times Grok replied to an X user; and the number of times a user responded to a Grok post. This list also included a second version of these queries restricted to only English-language posts. From September 2 to 8, we repeated these queries in the 11 languages with the most Grok calls after English, based on an initial sample of counts from all languages in the X API documentation.

\vspace{2mm}
\noindent
\myparagraph{Conversations Collected by Hour}
To collect interactions with Grok we used the \texttt{recent-search} X API endpoint, which returns a sample of posts matching a query (in our case, posts from Grok within a given time range). We then reconstructed tweet chains by following the reply hierarchy upward from these sampled posts. We sampled 500 Grok posts per day to fit within the quota of our X API subscription. To capture the empirically observable temporal variability of Grok interactions, we varied post collection across hours of the day. To determine how many posts to collect per hour, we monitored hourly post counts over the week prior to the start of the data collection and computed the average proportion of posts occurring in each hour of the day. Table~\ref{tab:convhour} shows the number and proportion of posts observed during that week, as well as the corresponding number of posts (out of 500) collected for each hour. This approach allows us to capture a random sample of posts throughout the day while accounting for natural ebbs and flows in activity.

\vspace{2mm}
\noindent
\myparagraph{Chain Conversation Reconstruction}
To obtain focal post chains, we first used the \texttt{recent-search} endpoint to request posts wherein Grok replied to a user, restricting results to English to allow for analysis by the authors. In the API's response, we also received the posts to which Grok replied, corresponding to the Grok Prompt posts in our focal post chains. We extracted the \texttt{id} of the Direct Parent post (the post to which the Grok Prompt post replies), as well as the \texttt{conversation\_id} (the post \texttt{id} of the Conversation Root). We filtered posts for which the \texttt{conversation\_id} was equal to Grok Prompt or Direct Parent \texttt{id}. We then retrieved Direct Parent and Conversation Root posts via the \texttt{tweet-lookup} endpoint. We designed our data collection to maximize ecological validity: if a user navigates to the Grok ``With Replies'' page on X (\url{https://x.com/grok/with_replies}), they see a series of ``profile conversations,'' each with 3-4 posts that end with a Grok reply, preceded a Grok prompt, similar to our chains.

\begin{table}[h!]
\centering
\footnotesize
\setlength{\tabcolsep}{4pt}
\begin{tabular}{l>{\color{black}}m{1.2cm} >{\color{black}}m{1.5cm} >{\color{black}}m{2.45cm}}
\toprule
Hour of the day &  Average Count of Posts & Average Proportion & Sample Target (500 posts per day) \\
\midrule
0  & 32,808 & 0.039 & 20 \\
1  & 31,581 & 0.037 & 18 \\
2  & 31,458 & 0.037 & 18 \\
3  & 31,957 & 0.038 & 19 \\
4  & 30,470 & 0.036 & 18 \\
5  & 29,398 & 0.035 & 18 \\
6  & 29,247 & 0.034 & 17 \\
7  & 29,058 & 0.034 & 17 \\
8  & 28,187 & 0.033 & 16 \\
9  & 27,622 & 0.033 & 16 \\
10 & 28,594 & 0.034 & 17 \\
11 & 29,672 & 0.035 & 18 \\
12 & 35,230 & 0.042 & 21 \\
13 & 36,710 & 0.043 & 22 \\
14 & 38,614 & 0.046 & 23 \\
15 & 40,898 & 0.048 & 24 \\
16 & 44,078 & 0.052 & 26 \\
17 & 45,860 & 0.054 & 27 \\
18 & 43,874 & 0.052 & 26 \\
19 & 47,488 & 0.056 & 28 \\
20 & 43,153 & 0.051 & 26 \\
21 & 41,652 & 0.049 & 24 \\
22 & 39,409 & 0.046 & 23 \\
23 & 31,022 & 0.037 & 18 \\
\bottomrule
\end{tabular}
\caption{Conversations Collected by Hour (UTC)}
\label{tab:convhour}
\end{table}

\newpage

\subsection{Language Distribution}
We include the distribution of languages for Grok prompt and Grok reply posts below in Figure~\ref{fig:language_distribution}. 

\begin{figure}[h!]
    \centering
    \includegraphics[width=0.9\linewidth]{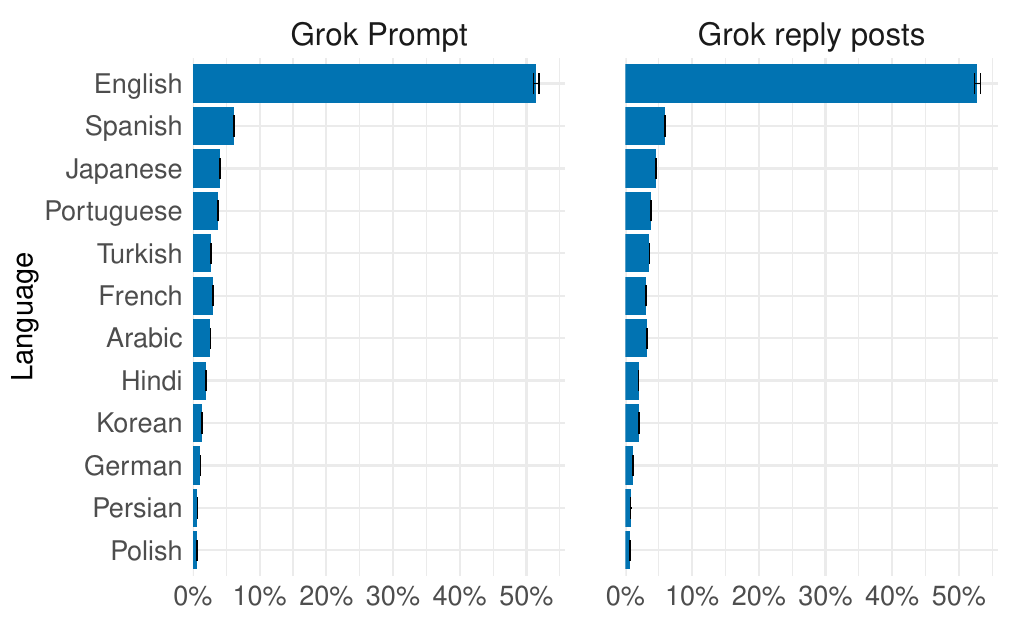}
    \caption{Language distribution of Grok prompt and Grok reply posts, 9/2-8, 2025. English is the predominant language among Grok prompt and reply posts, followed by Spanish and Portuguese. Error bars show standard errors.}
    \label{fig:language_distribution}
\end{figure}

\subsection{Engagement Analysis}
We include the analysis of the difference in engagement between the Conversation root and Grok reply posts in Table~\ref{tab:engagement_comparison}. 

\begin{table}[th]
\begin{tabular}{>{\color{black}}m{3cm}>{\color{black}}m{3cm}}
\toprule
Engagement Metric & D statistic\\
\midrule
Views  &  $0.73^{***}$\\
Likes  & $0.78^{***}$\\
Retweets  &  $0.77^{***}$\\
Replies & $0.77^{***}$\\
\bottomrule
\end{tabular}
\caption{\added{Engagement Comparison between Grok Reply and Root posts: results from two-sample Kolmogorov-Smirnov tests yield significant differences in all engagement metrics between Grok Reply and Root posts. Notation: ***indicates $p<0.001$.}}
\begin{minipage}{0.9\linewidth}
\caption*{\textit{Note:} \added{The Kolmogorov--Smirnov D statistic ranges from 0 to 1 and represents the maximum difference between empirical cumulative distribution functions. Larger values indicate greater divergence between the compared engagement distributions.}
}
\end{minipage}
\label{tab:engagement_comparison}
\end{table}

\subsection{Grok Contextual Analysis: Stratification Weights}
We include weights in Table~\ref{tab:sample_weight} used to stratify our sampling of Grok usage categories for contextual analysis.

\begin{table}[h!]
\centering
\footnotesize
\begin{tabular}{p{4.7cm}cc}
\toprule
\textbf{Index} & \textbf{\# Samples} & \textbf{Weight} \\
\midrule
General Information-Seeking        & 328 & 0.32 \\
Fact-checking                      & 128 & 0.13 \\
Support one's argument             & 120 & 0.12 \\
Opinion and Advisory Use           & 89 & 0.09 \\
Prompt Grok for Creative and Generative Interactions & 87 & 0.15 \\
Debate with Grok                   & 79 & 0.08 \\
Ask Grok information about X Platform or Users & 49 & 0.05 \\
Criticize Grok's Behaviors / Provide Negative Comments on Grok & 48 &0.05 \\
Positive Comments for Grok          &  29 & 0.03 \\
Asking Grok Questions About Itself and About AI  &  27 & 0.03 \\
Other                              &  18 & 0.02 \\
\bottomrule
\end{tabular}
\caption{Sample Count and Stratification Weight by Index Category}
\label{tab:sample_weight}
\end{table}

\subsection{\added{Breakdown of Categories of Interactions by Type of Focal Post Chains}}
\added{We include the breakdown of categories of interaction with Grok based on where Grok prompts occur within the conversation chains in Figure~\ref{fig:overall_classification_distribution_breakdown}.} 
\begin{figure*}[th!]
    \centering    \includegraphics[width=1\linewidth]{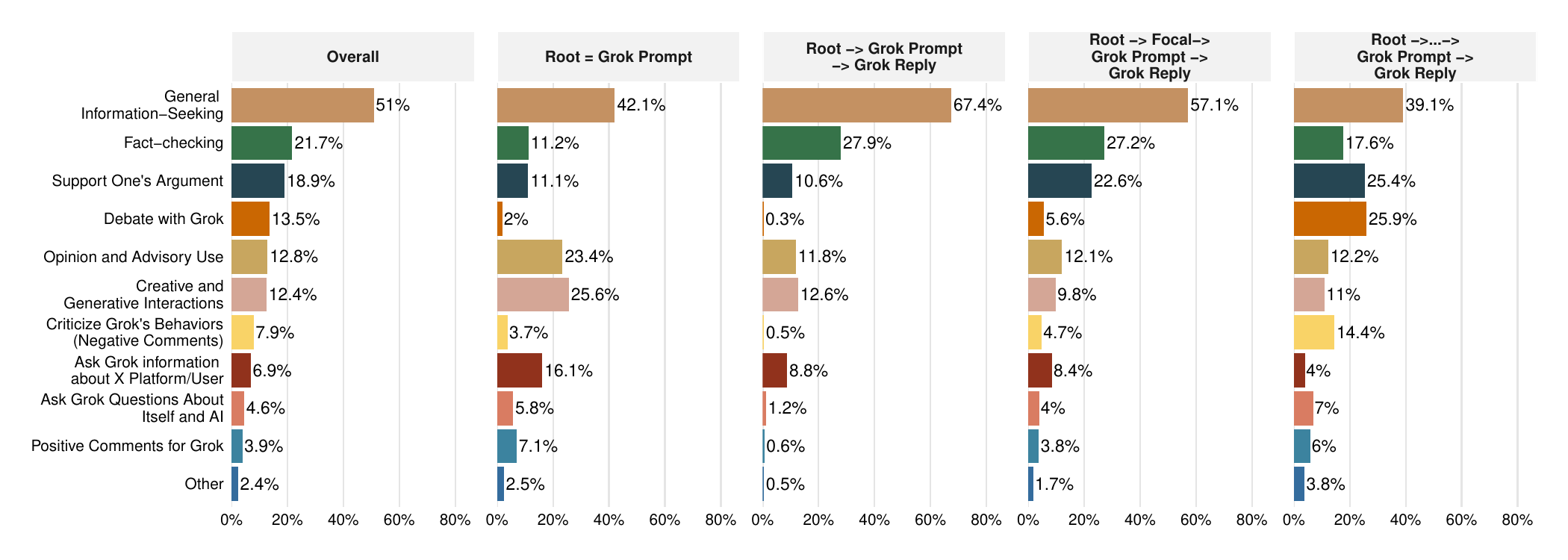}
    \caption{Categories of interaction with Grok, with proportion of focal post chains assigned to each category. Note that each interaction can be assigned to more than one category. Information seeking and fact-checking were the most common categories, and they occurred with greater frequency deeper in post chains.}
    \label{fig:overall_classification_distribution_breakdown}
\end{figure*}

\newpage

\subsection{LLM Prompt for Grok Use Classification}\label{app:llmprompt}

Below we include the complete prompt we provided to Google Gemini-Pro-2.5 for classifying the posts in our dataset according to the use categories we created via our initial content analysis. The prompt we used for classifying roles is included in the GitHub repository of the project due to its length.

\begin{tcolorbox}[colback=gray!5,
                  colframe=black!75,
                  boxrule=0.5pt,
                  arc=2mm,
                  left=2pt,
                  right=2pt,
                  top=2pt,
                  bottom=2pt,
                  width=\columnwidth,
                  enhanced,
                  sharp corners,
                  breakable]
Below is a tweet thread that consists of root tweet, focal tweet, grok call, and grok reply. 

\begin{itemize}
    \item Grok call is a tweet where the user calls grok for response. 
    \item Grok reply is the response from grok to the grok call.
\end{itemize}

Please provide the overall topic of the conversation, e.g., politics, technology, health, entertainment, sports, etc.  

Then describe the type of information that grok provides or generates in grok's reply, e.g., factual information, opinion, creative content, statistics about movies.  

Then provide a brief description of what the user is requesting from grok, e.g., asking for factual information about political event, asking for grok's opinion on a sports event.  

Then classify the user's request to Grok into one or more of the following categories. It may belong to multiple categories.  

\begin{enumerate}
  \item General Information-Seeking 
     \begin{itemize}
         \item User requests factual information, explanations, or knowledge on various topics.
         \item Excluding any requests that aim to seek opinions or verify information.
         \item Excluding any requests that ask grok to endorse their own argument or statement.
         \item If users are asking information about X platform or X users, select category 9. Ask Grok information about X Platform or Users.
     \end{itemize}
  \item Opinion and Advisory Use
     \begin{itemize}
         \item User requests Grok's opinion, suggestions, recommendations, or predictions on any matter.
     \end{itemize}
  \item Fact-checking
     \begin{itemize}
         \item User asks Grok to verify the accuracy of information, claims, or statements.
     \end{itemize}
  \item Support one's argument
     \begin{itemize}
         \item User asks Grok to endorse their statement or argument in the conversation.
         \item For example, user ask grok: ``Am I right?''; ``Tell them this is wrong.''
     \end{itemize}
  \item Debate with Grok
     \begin{itemize}
         \item Users challenge, question, or debate Grok's logic, reasoning, behavior, or outputs.
     \end{itemize}
  \item Criticize Grok's Behaviors / Provide Negative Comments on Grok
     \begin{itemize}
         \item Users express criticism, dissatisfaction, or negative attitudes toward Grok's responses or capabilities.
     \end{itemize}
  \item Positive Comments for Grok
     \begin{itemize}
         \item Users express appreciation, praise, or positive attitudes toward Grok's performance or helpfulness.
     \end{itemize}
  \item Prompt Grok for Creative and Generative Interactions
     \begin{itemize}
         \item User asks Grok to produce creative content, generate media, or analyze multimedia files, media content (images, videos, audio).
     \end{itemize}
  \item Ask Grok information about X Platform or Users
     \begin{itemize}
         \item Users ask Grok to query data or statistics of visitors or users. 
         \item Users ask Grok to access, detect, or analyze information about people, accounts, or content.
     \end{itemize}
  \item Asking Grok Questions About Itself and About AI
     \begin{itemize}
         \item Probing whether Grok has emotion, consciousness, or sentience.
         \item Probing whether Grok is objective or influenced by corporate interests.
     \end{itemize}
  \item Other
     \begin{itemize}
         \item Interactions that don't fit clearly into the above categories or are ambiguous.
     \end{itemize}
\end{enumerate}

Here is the tweet conversation:  

\begin{quote}
\{tweet\_conversation\}
\end{quote}

Please provide your answer in this format and do not need to include description which is after each colon of the category:  

\begin{quote}
\textbf{Overall topic of the conversation}: \texttt{<}overall topic of the conversation\texttt{>} \\
\textbf{Type of information grok provides}: \texttt{<}brief description of the type of information grok provides in grok's reply\texttt{>} \\
\textbf{User Request}: \texttt{<}brief description of what the user is requesting from grok\texttt{>} \\
\textbf{Category}: \texttt{<}name of the category or a comma-separated list of category names if there are more than one category\texttt{>}
\end{quote}
\end{tcolorbox}

\newpage

\subsection{Topic Model}
We include the visualization of topics discussed with Grok by X users in Figure~\ref{fig:content_topic_map_primary}. In Table~\ref{tab:topics}, we include the sizes (in conversation count) of the clusters produced for our topic model of user conversations with Grok. \added{We additionally include a complementary topic model with more fine-grained topics in Figure~\ref{fig:content_topic_model_secondary}.} In Table~\ref{tab:user_topic_freq}, we describe the sizes (in user bio count) of clusters produced for our topic model of user bios.
\added{Similarly, we included a complementary topic model for user bios in Figure~\ref{fig:user_topic_model_secondary}.}

\begin{table}[!h]
\centering
\begin{tabular}{p{6cm}r}
\toprule
\textbf{Topic Label} & \textbf{Topic Size} \\
\midrule
Uncategorized & 8795\\

Social issues and identity & 7168\\

Social Media Engagement & 1695\\

Israel-Palestine Conflict & 1532\\

Indian Politics Dynamics & 1409\\

Football transfers news & 1362\\

Online Community Posts & 1142\\

Trump and FBI allegations & 964\\

Islam and Christianity Conflict & 916\\

Migration and politics & 879\\

Online conversation threads & 858\\

AI and Elon Musk & 801\\

Charlie Kirk assassination & 630\\

Russia-Ukraine Conflict & 609\\

Government Shutdown Impact & 506\\

Nigerian politics and governance & 436\\

Elon Musk Innovations & 399\\

Advances in AI & 372\\

Immigration enforcement actions & 367\\

Crime and Accountability & 353\\

Interstellar phenomena & 351\\

Sports Betting Insights & 346\\

Sports News Highlights & 324\\

Cryptocurrency market trends & 310\\

Vaccine controversies and safety & 302\\

Music Industry Trends & 282\\

Adult content themes & 276\\

Nigerian social dynamics & 265\\

Afghanistan-Pakistan conflict & 263\\

Kenyan political landscape & 237\\

Tariffs and Economy & 231\\

South Africa's Political Climate & 226\\

Canadian political issues & 221\\

Indonesia political unrest & 218\\

Pakistan cricket updates & 209\\

Sexual relationships and drama & 207\\

Stock Market Trends & 207\\

Indian Film Industry & 202\\

Creative Animation Projects & 198\\

Federal Reserve Actions & 189\\

Indian social issues & 176\\

Transgender discourse and controversies & 174\\

NYC mayoral election & 165\\

Trump and Epstein Files & 162\\

Gavin Newsom controversies & 150\\

Customer service complaints & 145\\

AI Art Generation & 144\\

Twitter personality analysis & 133\\

Gun violence debate & 132\\

Social media interactions & 122\\

U.S. Military Actions in Venezuela & 117\\

Wataa references & 113\\

Energy policy debates & 108\\
\bottomrule
\end{tabular}
\caption{Cluster Sizes for Topic Model of Posts via BERTTopic modeling.}
\label{tab:topics}
\end{table}

\begin{figure*}[!h]
    \centering    
    \includegraphics[width=0.8\linewidth]{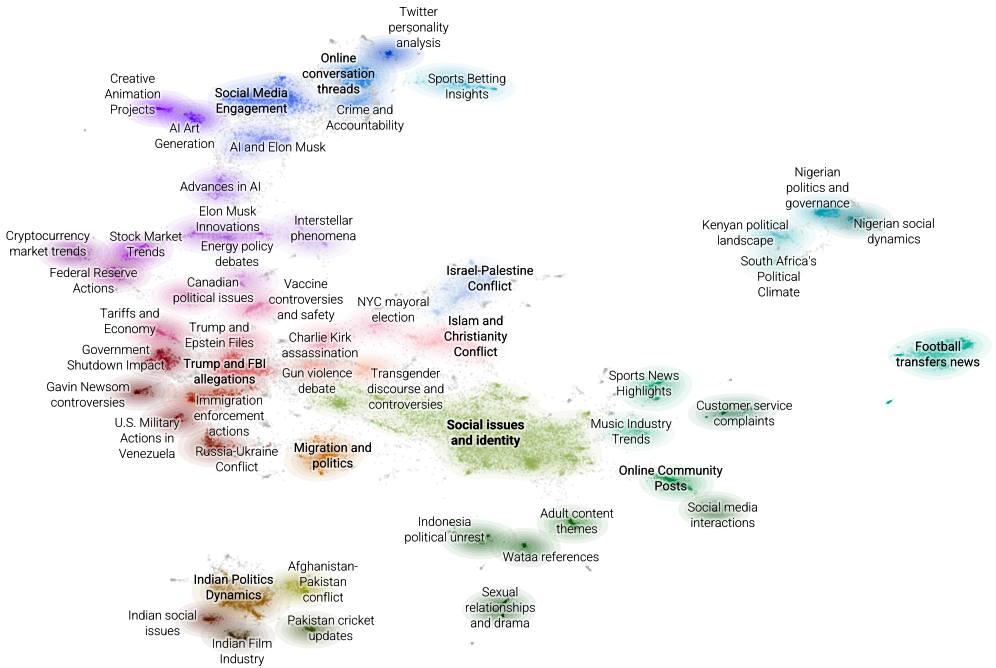}
    \caption{A topic model depicting the domains of topics discussed with Grok by X users, reflecting the centrality of the political, cultural, and technological topics that shape Grok's social roles and categories of use.} 
    \label{fig:content_topic_map_primary}
\end{figure*}
\begin{figure*}[!h]
    \centering    \includegraphics[width=0.9\linewidth]{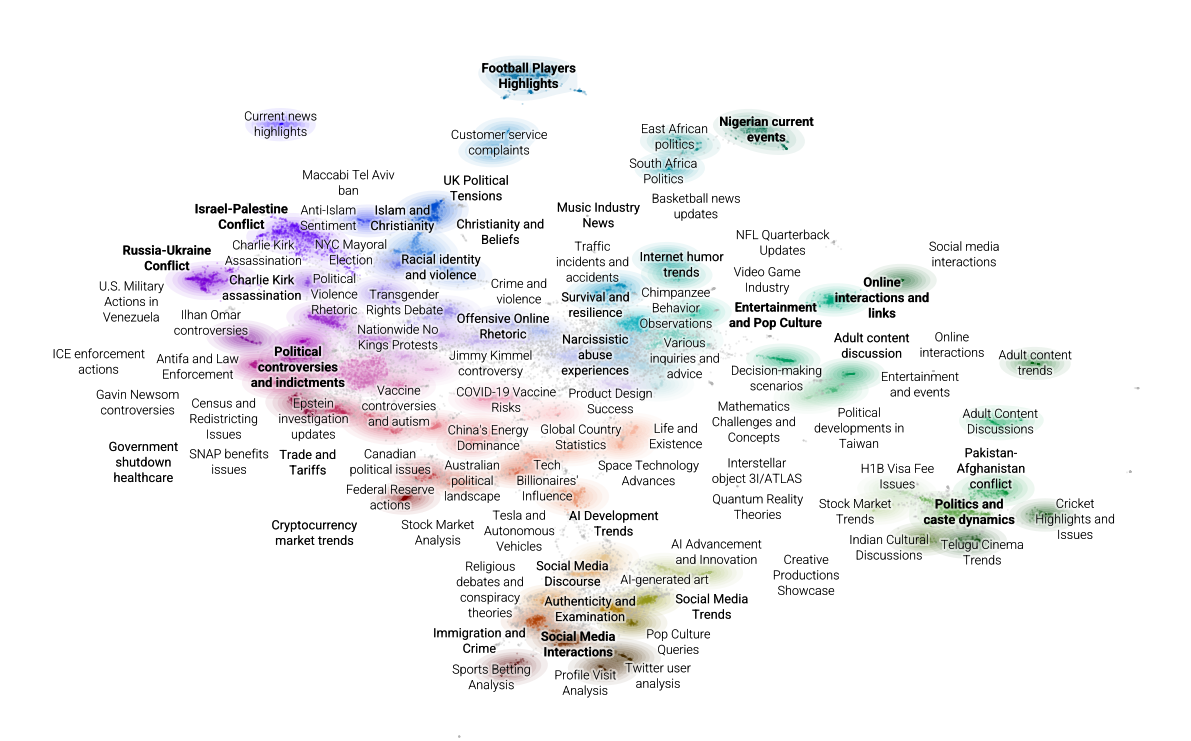}
    \caption{Complementary topic model with more fine-grained domains of topics discussed with Grok by X users. We set the minimum cluster size to be 50 examples and reduced dimensionality using UMAP with 20 neighbors. }
    \label{fig:content_topic_model_secondary}
\end{figure*}

\begin{figure*}[!t]
    \centering    
\includegraphics[width=0.8\linewidth]{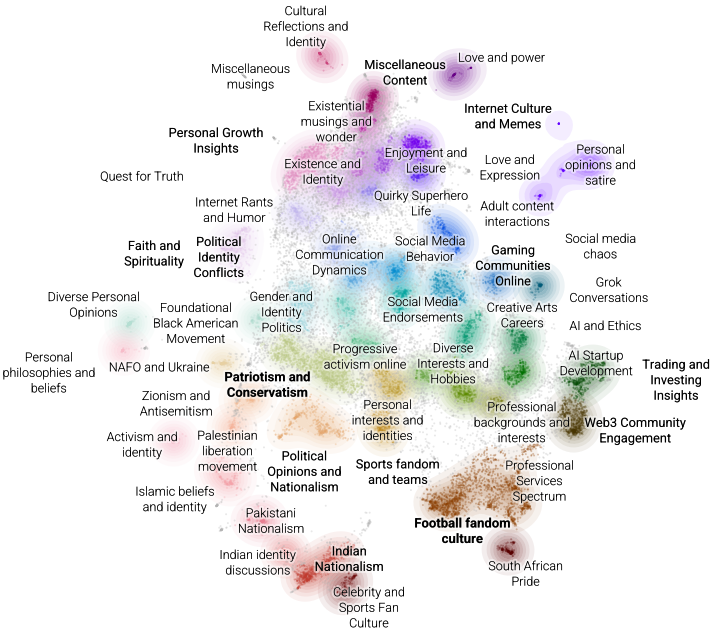}
    \caption{Topic model of \newdatainfo{22,884} user profile descriptions: top 3 by size are bolded: ``Football fandom culture,'' ``Patriotism and Conservatism,'' and ``Indian Nationalism.''}    \label{fig:user_topic_model}
\end{figure*}

\begin{figure*}[!th]
    \centering    
\includegraphics[width=0.8\linewidth]{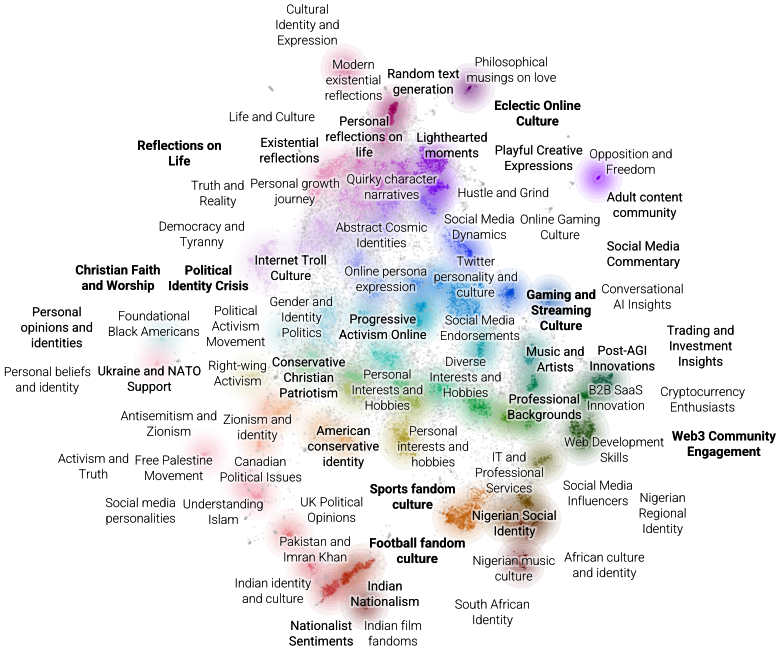}
    \caption{Complementary topic model with more fine-grained user clusters. We set the minimum cluster size to be 20 examples and reduced dimensionality using UMAP with 20 neighbors. }    \label{fig:user_topic_model_secondary}
\end{figure*}

\newpage
\subsection{\added{Description of Grok Roles, and Breakdown of Roles by User Cluster}}
\added{We include the descriptions of roles of Grok assumed by X users in Table~\ref{tab:grok-roles}. We include a detailed breakdown of roles by the top 20 user clusters in Figure~\ref{fig:cross_user_cluster_roles} and Figure~\ref{fig:cross_user_cluster_roles_full}. Note that each percentage represents the proportion of users in each cluster who assumed Grok to take a particular role.}

\begin{table}[!t]
\centering

\begin{tabular}{p{6cm}r}
\toprule
\textbf{Keywords} & \textbf{Topic Size} \\
\midrule
Uncategorized & 9369\\
Football fandom culture & 1718\\

Patriotism and Conservatism & 1334\\

Indian Nationalism & 716\\

Personal Growth Insights & 649\\

Web3 Community Engagement & 604\\

Faith and Spirituality & 480\\

Trading and Investing Insights & 463\\

Internet Culture and Memes & 444\\

Political Opinions and Nationalism & 437\\

Political Identity Conflicts & 391\\

Sports fandom and teams & 312\\

Gaming Communities Online & 310\\

Miscellaneous Content & 265\\

Internet Rants and Humor & 253\\

AI Startup Development & 243\\

Professional backgrounds and interests & 231\\

Social media chaos & 230\\

Creative Arts Careers & 227\\

Islamic beliefs and identity & 212\\

AI and Ethics & 202\\

Social Media Endorsements & 194\\

Enjoyment and Leisure & 188\\

Adult content interactions & 182\\

Personal interests and identities & 181\\

Online Communication Dynamics & 176\\

Celebrity and Sports Fan Culture & 169\\

Quest for Truth & 158\\

Personal opinions and satire & 153\\

Existence and Identity & 152\\

Diverse Personal Opinions & 144\\

Love and Expression & 139\\

South African Pride & 136\\

Existential musings and wonder & 131\\

Pakistani Nationalism & 129\\

Social Media Behavior & 128\\

NAFO and Ukraine & 127\\

Zionism and Antisemitism & 127\\

Progressive activism online & 123\\

Palestinian liberation movement & 108\\

Gender and Identity Politics & 104\\

Professional Services Spectrum & 104\\

Miscellaneous musings & 100\\

Diverse Interests and Hobbies & 86\\

Foundational Black American Movement & 79\\

Love and power & 76\\

Cultural Reflections and Identity & 75\\

Quirky Superhero Life & 73\\

Grok Conversations & 71\\

Activism and identity & 68\\

Personal philosophies and beliefs & 59\\

Indian identity discussions & 54\\

\bottomrule
\end{tabular}
\caption{Topic Frequency for User Profile Information}
\label{tab:user_topic_freq}
\end{table}

\begin{table*}[t]
\centering

\small
\begin{tabularx}{\textwidth}{p{2.2cm} p{4.5cm} X}
\toprule
\textbf{Role} & \textbf{Description} & \textbf{Examples} \\
\midrule
Oracle & Users approach Grok as if it can answer nearly any question; serves as a default role for information retrieval & Identifying details in media (song names, t-shirt brands); explaining scientific phenomena; translation; summarization; underspecified queries like ``help me out here'' \\
\addlinespace
Advisor & Users request advice and opinions, treating the model as an authority figure & Ethics questions (dating coworkers, military support); everyday advice (picnic spots, party RSVPs, fantasy football); predictions about future events (vaccine mandates); expert commentary on acquisitions \\
\addlinespace
Truth Arbiter & Users invoke Grok to decide veracity of disputed claims and dispel misinformation & Judging truth of public figure statements; verifying rumors (e.g., Trump death rumor); determining if images are AI-generated; evaluating contested claims about ``white genocide,'' women in science, colonialism \\
\addlinespace
Advocate & Users seek validation of their information and opinions, often in disagreements with others & Phrases like ``isn't it the case that'' or ``am I not right''; quoting prior Grok posts as evidence; iteratively building arguments with Grok; user prompting Grok to list COVID anti-vaxxers who died \\
\addlinespace
Adversary & Users disagree with Grok, and the model takes opposing positions, leading to debates; notably this role exhibits little sycophancy and is marked by `pushing-back' on perceived biases & Grok aligns with one user against another; users correct model biases/inaccuracies; civil debate with Grok as antagonist; Grok defends responses with ``I stand by the facts'' \\
\addlinespace
Stooge & Users attack Grok as flawed or controlled by external forces & Accusations of left-wing bias; claims of capture by mainstream media; allegations of censorship by X; criticism of verbosity and multimodal parsing failures; claims Grok increases misinformation \\
\addlinespace
Platform Insider & Users request inside information about X platform and its users & Summarizing perspectives across X userbase; explaining Community Notes decline; finding information about specific users; account analytics (top visitors); censorship/moderation inquiries; botnet analysis \\
\addlinespace
Platform Guardian & Users acknowledge or express appreciation for Grok's positive impacts on X & Praising Grok for correcting misinformation; testing and validating Grok's accuracy; requesting reproaches of users engaged in hate speech \\
\addlinespace
Creative Partner & Users engage Grok in creative or artistic expression & Developing recipes; creating emoji ciphers; satirical guides; personality assessments from X history; writing in style of public figures; extended text-based simulations and roleplay \\
\addlinespace
Research Subject & Users treat Grok as a topic of curiosity itself; seeking to understand how Grok operates, and the guardrails imposed by X & Questions about consciousness, gender, belief in God; probing capabilities (video generation, financial advice); reconciling inconsistencies; asking about Grok's relationship to Elon Musk \\
\bottomrule
\end{tabularx}
\caption{\added{Taxonomy of Roles in Human-Grok Interactions.}}

\label{tab:grok-roles}
\end{table*}
\begin{figure*}[!h]
    \centering
    \includegraphics[width=1\linewidth]{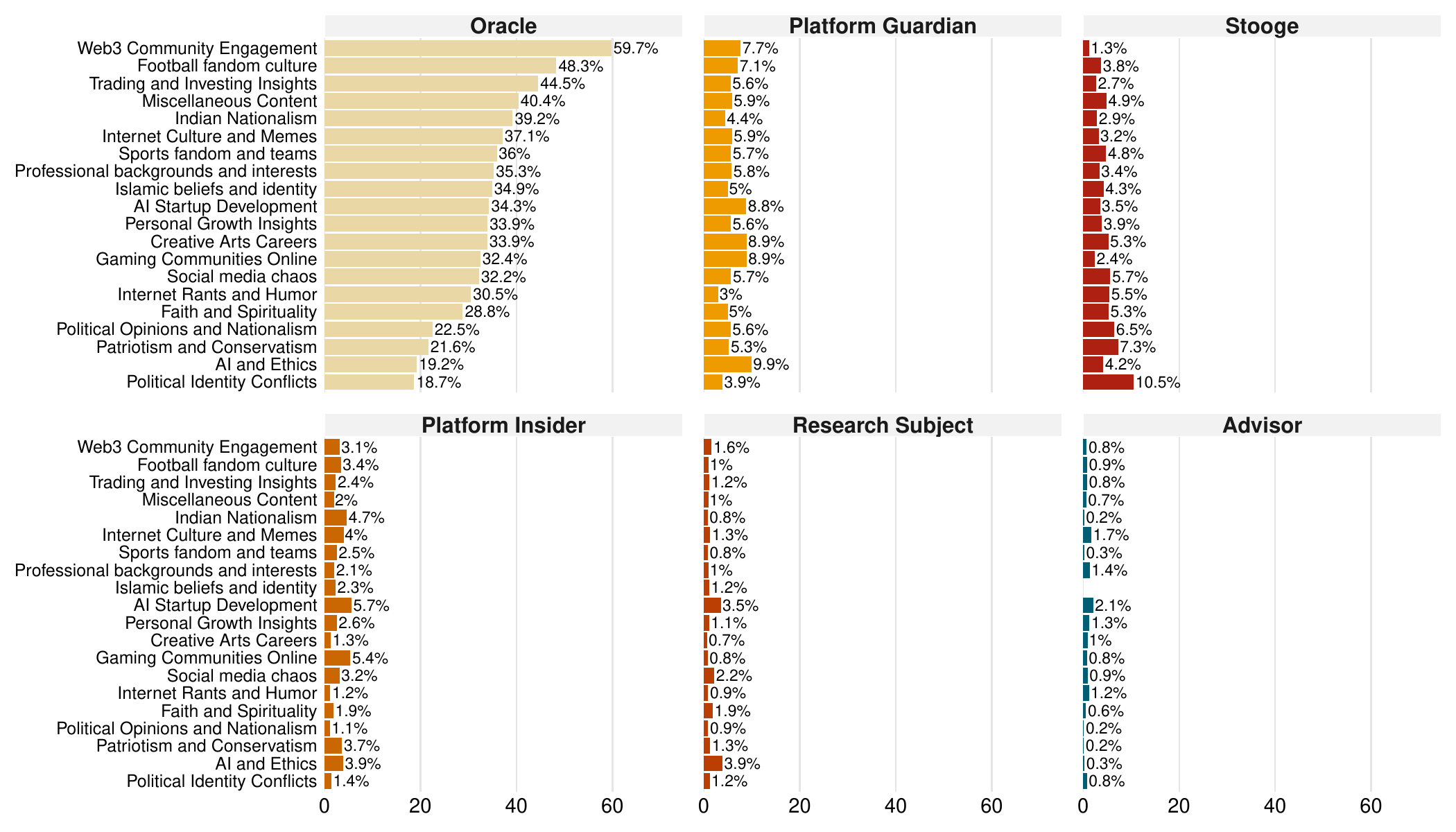}
    \caption{\added{Proportion of Grok roles within the twenty largest user clusters. 
    }} 
    \label{fig:cross_user_cluster_roles}
\end{figure*}

\begin{figure*}[!h]
    \centering
    \includegraphics[width=1\linewidth]{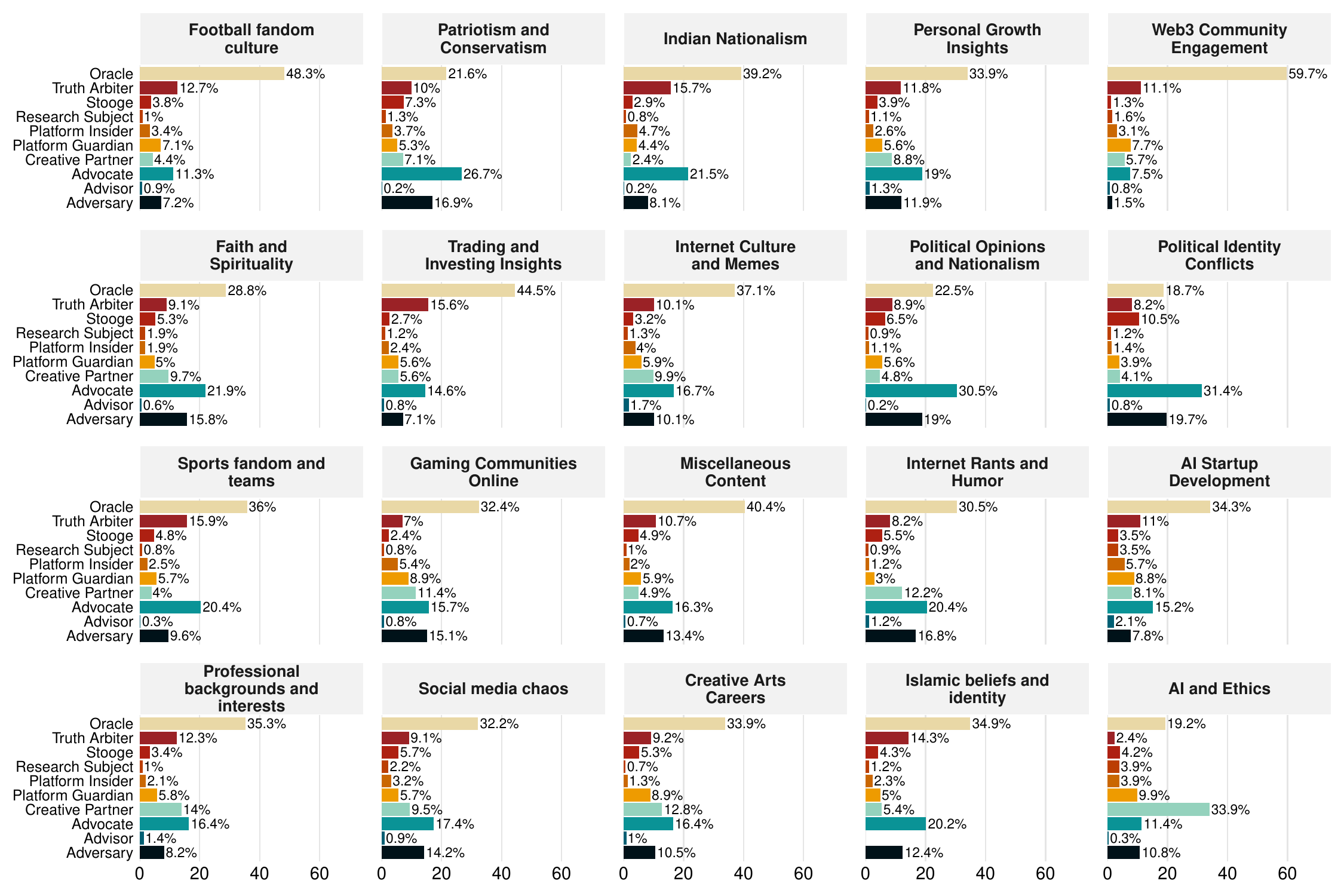}
    \caption{\added{Breakdown of Grok social role within the top twenty largest user clusters.}}
    \label{fig:cross_user_cluster_roles_full}
\end{figure*}

\end{document}